\documentclass[aps,pra,twocolumn,notitlepage,superscriptaddress,10pt,floatfix]{revtex4-1}
\usepackage[pdftex,plainpages=false,colorlinks=true,citecolor=blue,linkcolor=blue,urlcolor=blue,filecolor=green,bookmarksopen=true]{hyperref}
\usepackage[utf8]{inputenc}
\usepackage{amsmath}
\usepackage[caption = false]{subfig}
\usepackage{graphicx,epstopdf}
\usepackage{blindtext}
\usepackage{lipsum}
\usepackage{amsfonts}
\usepackage{bbm}
\usepackage{amssymb}
\usepackage{enumerate}
\usepackage{color}
\usepackage{latexsym}
\usepackage{times,txfonts}
\newtheorem{theorem}{Theorem}

\newtheorem{definition}[theorem]{Definition}

\newcommand{\Cmath}{\mathcal{C}}

\newcommand{\Hcal}{\mathcal{H}}
\newcommand{\Gcal}{\mathcal{G}}
\newcommand{\Dcal}{\mathcal{D}}

\newcommand{\Acal}{\mathcal{A}}
\newcommand{\Lcal}{\mathcal{L}}
\newcommand{\Ecal}{\mathcal{E}}

\newcommand{\Ocal}{\mathcal{O}}
\newcommand{\Ical}{\mathcal{I}}
\newcommand{\Fcal}{\mathcal{F}}
\newcommand{\Ucal}{\mathcal{V}}

\newcommand{\1}{\mathbbm{1}}
\newcommand{\Lmath}{\mathbbm{L}}

\newcommand{\dket}[1]{| #1 \rangle\!\rangle}
\newcommand{\dbra}[1]{\langle\!\langle #1 |}
\newcommand{\ket}[1]{| #1 \rangle}

\newcommand{\dinterpro}[2]{\langle\!\langle #1 | #2 \rangle\!\rangle}

\newcommand{\bra}[1]{\langle #1 |}
\newcommand{\tr}[1]{ \text{Tr}\left\{ #1 \right\}}
\newcommand{\trs}[1]{ \text{Tr} \{ #1 \}}

\begin{document}

\title{Sufficient conditions for adiabaticity in open quantum systems}

\author{Alan C. Santos}
\email{ac\_santos@df.ufscar.br}
\affiliation{Departamento de F\'{i}sica, Universidade Federal de S\~ao Carlos, P.O. Box 676, 13565-905, São Carlos, São Paulo, Brazil}

\author{Marcelo S. Sarandy}
\email{msarandy@id.uff.br}
\affiliation{Instituto de F\'{i}sica, Universidade Federal Fluminense, Av. Gal. Milton Tavares de Souza s/n, Gragoat\'{a}, 24210-346 Niter\'{o}i, Rio de Janeiro, Brazil}

\begin{abstract}
The adiabatic approximation exhibits wide applicability in quantum mechanics, providing a simple approach for non-transitional dynamics in quantum systems governed by slowly 
varying time-dependent Hamiltonians. However, the standard adiabatic theorem is specifically derived for closed quantum systems. 
In a realistic open system scenario, the inevitable system-reservoir interaction must be taken into account, which strongly impacts the generalization of the adiabatic behavior. 
In this paper, we introduce new sufficient conditions for the adiabatic approximation in open quantum systems. These conditions are simple yet general, 
providing a suitable instrument to investigate adiabaticity for arbitrary initial mixed states evolving under time local master equations. 
We first illustrate our results by showing that the adiabatic approximation for open systems is compatible with the description of quantum thermodynamics at thermal equilibrium, where 
irreversible entropy production is vanishing. We also apply our sufficient conditions as a tool in quantum control, evaluating the adiabatic behavior for the Hamiltonians of both 
the Deutsch algorithm and the Landau-Zener model under decoherence. 
\end{abstract}

\maketitle

\section{Introduction}

Inverse quantum engineering is a useful approach to drive quantum systems through some desired path in parameter space and, 
hence, to achieve a target state~\cite{Chen:11,Jing:13,Kang:16,Yu:18,Santos:18-a,Chen:18-1}. Within a number of different approaches 
for inverse engineering, one can highlight the adiabatic dynamics~\cite{Born:28,Messiah:Book} as an important strategy, with successful 
applications in quantum thermodynamics~\cite{Kieu:04,Alicki:18,Hu:20-a,Deffner-Campbel:Book}, quantum control~\cite{Kral:07,Odelin:19}, 
and quantum computation~\cite{Farhi:01,Tameem:18}. However, the standard adiabatic theorem is specifically derived for closed quantum 
systems. In a real physical scenario, where the quantum system is coupled with a surrounding environment, the concept of adiabaticity  
requires a reformulation so that it may be applicable to a non-unitary evolution. 
In this direction, Ref.~\cite{Sarandy:05-1} has introduced the adiabatic behavior of an open system by replacing the closed system  
picture of a decoupled evolution of the Hamiltonian eigenspaces with distinct energy eigenvalues for a decoupled evolution of 
Lindblad-Jordan eigenspaces with distinct eigenvalues of the Lindbladian superoperator. This notion of adiabaticity has been consistently 
applied in different scenarios, such as quantum computation~\cite{Sarandy:05-2}, geometric phases~\cite{sarandy:06}, eigenstate 
tracking of open quantum systems~\cite{Jing:16}, and quantum thermodynamics~\cite{Hu:20-a}.

The adiabatic approximation for open system has unraveled a competition between the time scale for adiabaticity, which typically requires long times, 
and the time scale for the decohering rates, which typically require short times, yielding a finite time adiabatic regime. This  
has been experimentally observed in Ref.~\cite{Steffen:03}. From the theoretical side, finite time adiabaticity emerges from a  
general adiabatic condition involving a set of integral expressions containing exponentials with real and imaginary contributions~\cite{Sarandy:05-1,Sarandy:05-2}. 
This holds for general initial mixed quantum states evolving under time local master equations. 

Alternatively, the adiabatic approximation in open quantum systems may be also  
introduced by different physically motivated approaches, such as state purification embedded into non-Hermitian dynamics ~\cite{Yi:07}, 
noiseless subsystem decomposition~\cite{Oreshkov:10}, weak coupling limit~\cite{Patrik:05}, and 
instantaneous steady state evolution of the Liouvillian~\cite{Venuti:16}. 
Adiabatic theorems for generators of contracting evolutions have also been proposed 
based on the notion of parallel transport in the manifold of instantaneous stationary states, both for 
gapped and gapless cases of the spectrum of the generator~\cite{Avron:12}.
 Notice then that adiabaticity in open systems has been established as a multifaceted concept, which leads to distinct and potentially complementary 
simplifying strategies to solve the open quantum dynamics.  
Here, we will keep the original multidimensional Jordan block approach of Ref.~\cite{Sarandy:05-1}. 
This allows for the representation of arbitrary initial mixed states, 
distributed in general superpositions of Jordan subspaces, evolving under arbitrary time local evolution. 
Even though single Jordan blocks usually do not have individual physical interpretation, there are plenty of physical 
states that require superpositions of basis vectors belonging to different Jordan blocks to be represented. Indeed, we will 
provide examples of mixed states evolving under decoherence that require, from the beginning of the evolution, superpositions of distinct Jordan 
subspaces. Our examples will be based on the Lindblad superoperators for 
the Deutsch algorithm and for the Landau-Zener model. 

In a general setting,    
we will be interested in obtaining operational sufficient conditions for the adiabatic behavior.
More specifically, we aim at simplifying the original conditions in Ref.~\cite{Sarandy:05-1} but keeping them applicable in a general convolutionless dynamics. 
The new conditions are achieved through a derivation that can be interpreted as a generalization for open systems of the results obtained 
by D. M. Tong \textit{et al.}~\cite{Tong:07} for closed systems. We will analytically obtain two simultaneously required conditions, 
one of them yielding a standard gap condition, while the other simplifying the integral term usually dealt with the Riemann-Lebesgue lemma in the closed case (see, e.g., Ref.~\cite{Sarandy:04}). 
Both conditions can be compactly written in terms of the gaps in the Liouvillian spectrum.   

The manuscript is organized as follows. In Sec.~\ref{RevAdiabOS} we derive the main result of this paper, with the new conditions presented in Subsec.~\ref{SecAdCond}. 
In Subsec.~\ref{SecU}, we introduce the adiabatic open system evolution operator, which turns out to be a useful tool in the applications. In Sec.~\ref{SecApplications}, we then 
illustrate our results, first by discussing the relationship between the open system adiabatic approximation and quantum thermodynamics at thermal equilibrium, and then  by 
evaluating the adiabatic decohering dynamics for the Hamiltonians of both the Deutsch algorithm and the Landau-Zener model. 
In Sec.~\ref{SecConc}, we present our conclusions.

\section{Adiabatic dynamics in open quantum systems} \label{RevAdiabOS}

In this section, we will derive new sufficient conditions for the adiabatic approximation in open systems. As an initial step,  
we will revisit the adiabatic approximation in open quantum systems. This will be performed from the point of view of an open system evolution operator, 
which will be introduced as an intermediate by-product of this work. However, let us first discuss the mathematical framework of open systems in the superoperator 
formalism. We consider a quantum system described by a density operator $\rho(t)$ acting on a $D_{\text{S}}$-dimensional Hilbert space, whose evolution is governed by a time-local master equation
\begin{eqnarray}
\dot{\rho}(t) = \Lcal_{t}[\rho(t)] \text{ , } \label{Lcal}
\end{eqnarray}
where $\Lcal_{t}[\bullet]$ is a time-dependent dynamical generator and the overdot denotes time derivative. Here, we do not need to assume a particular $\Lcal_{t}[\bullet]$, but later on we will consider it 
in the Lindblad form
\begin{equation}
\Lcal_{t}[\bullet] = \frac{1}{i\hbar} [H(t),\bullet] + \frac{1}{2}\sum_{n} \left(2\Gamma_{n}(t)\bullet\Gamma^{\dagger}_{n}(t) - \left\{\Gamma^{\dagger}_{n}(t)\Gamma_{n}(t) , \bullet\right\}\right) \text{, } \label{EqLindForm}
\end{equation}
where $\Gamma_{n}(t)$ are the time-dependent Lindblad operators that describe the coupling between our system and the environment. 
Differently from the closed system case, we need now take into account the reservoir influence. In this scenario, a convenient approach is the superoperator formalism~\cite{Sarandy:05-1,Horn:Book}. 
To this end, we define a matrix basis composed by $D_{\text{S}}\times D_{\text{S}}$ matrices $\sigma_{n}$ in which $\trs{\sigma_{n} \sigma_{m}} = D_{\text{S}}\delta_{nm}$. 
In this formalism, Eq.~\eqref{Lcal} is rewritten as (see Appendix~\ref{ApenSupForm})
\begin{eqnarray}
\dket{\dot{\rho}(t)} = \Lmath (t) \dket{\rho(t)} \text{ , } \label{EqEqSuperLindEq}
\end{eqnarray}
where $\dket{\rho(t)}$ is a $D_{\text{S}}^2$-dimensional ``coherence" vector in Hilbert-Schmidt space~\cite{Muynck:Book}, 
whose components are $\varrho_{n}(t) = \trs{\rho(t)\sigma_{n}^{\dagger}}$. 
We define the $D_{\text{S}}^2\times D_{\text{S}}^2$-dimensional superoperator $\Lmath(t)$ through its matrix representation, with matrix elements provided by 
$\Lmath_{ki}(t)=(1/D_{S})\trs{\sigma_{k}^{\dagger} \Lcal [ \sigma_{i} ]}$. The inner product between two coherence vectors associated with density operators 
$\xi_{1}$ and $\xi_{2}$ is given by $\dinterpro{\xi_{1}}{\xi_{2}} = (1/D_{S})\trs{\xi^{\dagger}_{1}\xi_{2}}$, where the conjugate coherence vector $\dbra{\xi_{1}}$ has 
components given by $\trs{\xi^{\dagger}_{1}\sigma_{n}}$. In particular, for a two-level system, the Pauli basis $\Ocal_{\text{tls}} = \{\1, \sigma_{x} , \sigma_{y} , \sigma_{z} \}$ 
is a convenient choice, but we can adopt more sophisticated bases depending on the application~\cite{Santos:20c}.  

In general, due to the non-Hermiticity of $\Lcal[\bullet]$, the superoperator $\Lmath(t)$ is not diagonalizable. Then, the notion of adiabaticity used in closed systems cannot be 
directly applied here~\cite{Sarandy:05-1}. On the other hand, general operators can be rewritten in the Jordan canonical form, where $\Lmath(t)$ is given in a block-diagonal 
structure $\Lmath_{\text{J}}(t)$ with Jordan blocks $J_{\alpha}(t)$ associated with different time-dependent non-crossing eigenvalues $\lambda_{\alpha}(t)$ of $\Lmath(t)$~\cite{Horn:Book}. 
The Jordan form of $\Lmath(t)$ is obtained by a similarity transformation through a matrix $S(t)$, reading
\begin{eqnarray}
\Lmath_{\text{J}}(t) = S^{-1}(t) \Lmath(t) S(t) = \text{diag}\begin{bmatrix}J_{0}(t) & J_{2}(t) & \cdots & J_{N-1}(t)\end{bmatrix} \text{, } \,\,\,\,\,\,\,\,\,
\end{eqnarray}
where $N$ is the sum of the geometric multiplicities of all the eigenvalues $\lambda_\alpha$ and each block $J_{\alpha}(t)$ is given by
\begin{eqnarray}
J_{\alpha}(t) = 
\begin{bmatrix}
\lambda_{\alpha}(t) & 1   & 0        & \cdots & 0 \\
0 &\lambda_{\alpha}(t) & 1 & \cdots & 0 \\
\vdots & \ddots & \ddots & \ddots & \vdots \\
0 & \cdots & 0 & \lambda_{\alpha}(t) & 1  \\
0 & \cdots & \cdots & 0 & \lambda_{\alpha}(t)  
\end{bmatrix} \text{ . } \label{JDiago}
\end{eqnarray}
As an immediate consequence of the above $\Lmath_{\text{J}}(t)$ structure, we can see that $\Lmath(t)$ does not necessarily admit the existence of a basis of eigenvectors. 
Instead, we define \textit{right} $\dket{\Dcal_{\alpha}^{n_{\alpha}}(t)}$ and \textit{left} quasi-eigenvectors $\dbra{\Ecal_{\alpha}^{n_{\alpha}}(t)}$ of $\Lmath(t)$ 
associated with the Jordan block $J_{\alpha}(t)$, which are defined by
\begin{subequations}
\label{EqEqEigenStateL}
\begin{align}
\Lmath(t)\dket{\Dcal_{\alpha}^{n_{\alpha}}(t)} &= \dket{\Dcal_{n}^{(n_{\alpha}-1)}(t)} + \lambda_{\alpha}(t)\dket{\Dcal_{\alpha}^{n_{\alpha}}(t)} \text{ , } \label{EqEqEigenStateLa} \\
\dbra{\Ecal_{\alpha}^{n_{\alpha}}(t)}\Lmath(t) &= \dbra{\Ecal_{n}^{(n_{\alpha}+1)}(t)} + \dbra{\Ecal_{\alpha}^{n_{\alpha}}(t)}\lambda_{\alpha}(t) \text{ , } \label{EqEqEigenStateLb}
\end{align}
\end{subequations}
where $n_\alpha=1,\cdots,N_\alpha$, with $N_\alpha$ denoting the dimension of $J_\alpha$ and 
$\dket{\Dcal_{\alpha}^{(0)}\!(t)}$ and $\dbra{\Ecal_{\alpha}^{(N_{\alpha}+1)}\!(t)}$ denoting vanishing vectors.
The sets $\{\dket{\Dcal_{\alpha}^{n_{\alpha}}(t)}\}$ and $\{\dbra{\Ecal_{\alpha}^{n_{\alpha}}(t)}\}$ satisfy the bi-orthonormalization condition 
$\dinterpro{\Ecal_{m}^{\beta}(t)}{\Dcal_{n}^{\alpha}(t)} = \delta_{mn}\delta_{\beta\alpha}$. Thus, we can write the completeness relationship 
\begin{eqnarray}
\sum_{\alpha=0}^{N-1} \sum _{n_{\alpha} = 1}^{N_{\alpha}}  \dket{\Dcal_{\alpha}^{n_{\alpha}}(t)}\dbra{\Ecal_{\alpha}^{n_{\alpha}}(t)} = \1_{D_{\text{S}}^2\times D_{\text{S}}^2} \text{ , }
\end{eqnarray}
which holds for all $t\in [0,\infty)$.

\vspace{-0.27cm}

\subsection{Conditions for adiabaticity in open systems} \label{SecAdCond}

As discussed before, the fact that the superoperator $\Lmath(t)$ is not necessarily diagonalizable needs to be taken into account to define adiabaticity 
for a non-unitary evolution. In this work, we will adopt the general definition of adiabaticity as established in Ref.~\cite{Sarandy:05-1}. This is based on 
the Jordan decomposition of $\Lmath(t)$, which will be generically designed here as the Lindblad superoperator.

\begin{definition}[Adiabaticity in open systems]\label{DefAdiabOS}
	An open quantum system is said to undergo an adiabatic dynamics if the evolution of the density operator in its Hilbert-Schmidt space can be 
	decomposed into decoupled Lindblad-Jordan eigenspaces associated with distinct, time-dependent, non-crossing eigenvalues of $\Lmath(t)$.
\end{definition}

Thus, let us now derive under what conditions we can achieve the adiabatic behavior of an open quantum system. 
To this aim, let us to write the evolved state $\dket{\rho(t)}$ in the basis $\{\dket{\Dcal_{\alpha}^{n_{\alpha}}(t)}\}$ as
\begin{eqnarray}
\dket{\rho(t)} = \sum_{\alpha=0}^{N-1} \sum _{n_{\alpha} = 1}^{N_{\alpha}} r_{\alpha}^{n_{\alpha}}(t) \dket{\Dcal_{\alpha}^{n_{\alpha}}(t)} \text{ , } \label{EqRhoExpD}
\end{eqnarray}
where $r_{\alpha}^{n_{\alpha}}(t)$ are coefficients to be determined. By inserting Eq.~(\ref{EqRhoExpD}) in Eq.~\eqref{EqEqSuperLindEq} 
and using Eq.~\eqref{EqEqEigenStateL}, we obtain
\begin{eqnarray}
&&\dot{r}_{\beta}^{k}(t) 
 = \lambda_{\beta}(t)r_{\beta}^{k}(t) - r_{\beta}^{k}(t) \dinterpro{\Ecal_{\beta}^{k}(t)}{\dot{\Dcal}_{\beta}^{k}(t)} + r_{\beta}^{k+1}(t) \hspace{1.8cm}
 \nonumber \\ 
&&- \sum_{n_{\beta} \neq k}^{} r_{\beta}^{n_{\beta}}(t) \dinterpro{\Ecal_{\beta}^{k}(t)}{\dot{\Dcal}_{\beta}^{n_{\beta}}(t)} 
- \sum_{\alpha\neq \beta} \sum_{n_{\alpha}} r_{\alpha}^{n_{\alpha}}(t) \dinterpro{\Ecal_{\beta}^{k}(t)}{\dot{\Dcal}_{\alpha}^{n_{\alpha}}(t)},    \label{Eqrdot}
\end{eqnarray}
with $ r_{\beta}^{N_\beta+1}(t)\equiv 0$. 
The first two terms in the right-hand-side of the equation above are associated with perfect decoupled evolution. 
The remaining terms tell us about the coupling between the $k$-th vector in the block $\beta$ and all the other  basis vectors inside and outside $\beta$. 
Therefore, in agreement with the Definition~\ref{DefAdiabOS}, adiabaticity in the context of open systems requires 
to eliminate the last sum term in Eq.~(\ref{Eqrdot}), which promote transitions between Jordan blocks.

Before considering the most general case, let us first particularize our analysis to the case in which $\Lmath(t)$ 
admits a Jordan decomposition into one-dimensional Jordan blocks in Eq.~\eqref{JDiago}. 
Under this assumption, the quasi-eigenstate relations in Eqs.~\eqref{EqEqEigenStateL} become genuine eigenstate equations, which are given by
\begin{subequations}
	\label{EqEqEigenOneL}
	\begin{align}
	\Lmath(t)\dket{\Dcal_{\alpha}(t)} &= \lambda_{\alpha}(t)\dket{\Dcal_{\alpha}(t)} \text{ , } \label{EqEqEigenOneLa} \\
	\dbra{\Ecal_{\alpha}(t)}\Lmath(t) &= \dbra{\Ecal_{\alpha}(t)}\lambda_{\alpha}(t) \text{ . } \label{EqEqEigenOneLb}
	\end{align}
\end{subequations}
Hence, Eq.~\eqref{Eqrdot} can be reduced to
\begin{align}
\dot{r}_{\beta}(t) & = \lambda_{\beta}(t)r_{\beta}(t) - r_{\beta}(t) \dinterpro{\Ecal_{\beta}(t)}{\dot{\Dcal}_{\beta}(t)} \nonumber \\
& - \sum_{\alpha\neq \beta} r_{\alpha}(t) \dinterpro{\Ecal_{\beta}(t)}{\dot{\Dcal}_{\alpha}(t)} \text{ . } \label{EqrdotOne}
\end{align}
Now, we can define a new parameter $p_{\beta}(t)$ as
\begin{eqnarray}
r_{\beta}(t) = p_{\beta}(t) e^{\int_{t_{0}}^{t} \lambda_{\beta}(\xi)d\xi} \text{ , } \label{pDef}
\end{eqnarray}
so that, from Eq.~(\ref{EqrdotOne}), it follows that $p_{\beta}(t)$ is governed by
\begin{align}
\dot{p}_{\beta}(t) &= - \sum_{\alpha\neq \beta} p_{\alpha}(t) e^{\int_{t_{0}}^{t} \left[\lambda_{\alpha}(\xi) - \lambda_{\beta}(\xi)\right] d\xi}\dinterpro{\Ecal_{\beta}(t)}{\dot{\Dcal}_{\alpha}(t)} \nonumber \\
&- p_{\beta}(t) \dinterpro{\Ecal_{\beta}(t)}{\dot{\Dcal}_{\beta}(t)}\text{ . } \label{EqpdotOne}
\end{align}
The first term in the right-hand-side is responsible for the coupling of distinct Lindblad-Jordan eigenspaces during the evolution. 
If we are able to minimize its effects, we can approximate the dynamics to
\begin{eqnarray}
\dot{p}_{\beta}(t) \approx - p_{\beta}(t) \dinterpro{\Ecal_{\beta}(t)}{\dot{\Dcal}_{\beta}(t)} \text{ . }
\end{eqnarray}
Then, the adiabatic solution $r_{\beta}(t)$ for the dynamics can be immediately obtained from Eq.~\eqref{pDef}, reading
\begin{eqnarray}
r_{\beta}(t) = r_{\beta}(t_{0}) e^{\int_{t_{0}}^{t} \lambda_{\beta}(\xi)d\xi}
e^{- \int_{t_{0}}^{t} \dinterpro{\Ecal_{\beta}(\xi)}{\dot{\Dcal}_{\beta}(\xi)}d\xi} \text{ , } \label{Eqr1BJdec}
\end{eqnarray}
where we have used $p_{\beta}(t_{0}) = r_{\beta}(t_{0})$. In conclusion, if the system undergoes the adiabatic dynamics along a non-unitary process, the evolved state is
\begin{eqnarray}
\dket{\rho^{\text{1D}}_{\text{ad}}(t)} = \sum _{\alpha=0}^{N-1} r_{\alpha}(t_{0}) e^{\int_{t_{0}}^{t} \Lambda_{\alpha}(\xi)d\xi}\dket{\Dcal_{\alpha}(t)} \label{EqAdEvol1D} \text{ , }
\end{eqnarray}
with $\Lambda_{\alpha}(t)=\lambda_{\alpha}(t)-\dinterpro{\Ecal_{\alpha}(t)}{\dot{\Dcal}_{\alpha}(t)}$ being the generalized adiabatic phase accompanying the dynamics 
of the $n$-th eigenvector. Throughout this paper, the superscript ``1-D'' indicates that the result is valid by assuming that the Lindblad superoperator admits one-dimensional 
Jordan block decomposition. 

Conditions for the validity of the adiabatic dynamics can be properly derived by defining the normalized time $s=t/\tau$, with $\tau$ denoting the total evolution 
time and $0\le s \le 1$.  
For a one-dimensional Jordan decomposition of $\Lmath(t)$, a sufficient condition for the decoupled evolution of $\dket{\Dcal^{\text{DA}}_{\beta}(t)}$ from the 
remaining eigenvectors $\dket{\Dcal^{\text{DA}}_{\alpha\neq\beta}(t)}$, with $\lambda_\alpha \neq \lambda_\beta$, is provided by~(See Appendix~\ref{ApConditions})
\begin{subequations}\label{EqAdCondOpenSystem}
	\begin{align}
	\text{(C1)} ~~ & \left \vert \frac{\tilde{F}_{\alpha\beta}(s)  e^{\tau\int_{s_{0}}^{s} \Gcal_{\alpha\beta}(s^{\prime})ds^{\prime}}}{\tau\Gcal_{\alpha\beta}(s)} \right \vert \ll 1 \text{ , } \label{EqACOpenC1} \\
	\text{(C2)} ~~ &\left \vert \frac{1}{\tau}\int_{s_{0}}^{s} \frac{d}{ds^{\prime}} \left[ \frac{\tilde{F}_{\alpha\beta}(s^{\prime})}{\Gcal_{\alpha\beta}(s^{\prime})} \right]e^{\tau\int_{s_{0}}^{s^{\prime}} \Gcal_{\alpha\beta}(s^{\prime\prime}) ds^{\prime\prime}}
	ds^{\prime} \right \vert \ll 1 \text{ , } \label{EqACOpenC2}
	\end{align}
\end{subequations}
where we defined $\Gcal_{\alpha\beta}(s) = \lambda_{\alpha}(s) - \lambda_{\beta}(s)$ and 
\begin{equation}
\tilde{F}_{\alpha\beta}(s) = e^{-\int_{s_{0}}^{s} \dinterpro{\Ecal_{\beta}(s^{\prime})}{d_{s}\Dcal_{\beta}(s^{\prime})} ds^{\prime} } \dinterpro{\Ecal_{\beta}(s)}{d_{s}\Dcal_{\alpha}(s)} \text{ ,} \label{EqFalphaBeta}
\end{equation}
with $d_{s}f(s) \equiv df(s)/ds$. Conditions (C1) and (C2) are required to hold for every $s_0$ such that $s_0 \leq s \leq 1$. 
If they are satisfied for all $\alpha$, the $\beta$-th eigenvector evolves decoupled from the other eigenvectors such that $\lambda_\alpha \neq \lambda_\beta$. 
In case they are satisfied for 
all $\alpha$ and $\beta$, all eigenvectors of the spectrum of $\Lmath(t)$ evolve decoupled from each other. 
Moreover, it is worth highlighting here that the above conditions are very similar to the conditions as proposed by 
D. M. Tong \textit{et al.}~\cite{Tong:07} for closed systems. In fact, the second condition can be easily rewritten as
\begin{equation}
\text{(C2$^\prime$)} \quad \quad \left \vert \frac{1}{\tau} \frac{d}{ds^{\prime}} \left[ \frac{\tilde{F}_{\alpha\beta}(s^{\prime})}{\Gcal_{\alpha\beta}(s^{\prime})} \right]e^{\tau\int_{s_{0}}^{s^{\prime}} \Gcal_{\alpha\beta}(s^{\prime\prime}) ds^{\prime\prime}}
\right \vert_M  \ll 1 \text{ , } \label{EqACOpenC2Max}
\end{equation}
where we have used $\left \vert \int_{x_{0}}^{x_{1}}f(x)dx \right \vert \leq\!\int_{x_{0}}^{x_{1}} \left \vert f(x) \right \vert dx$, with the subscript $M$ in Eq.~(\ref{EqACOpenC2Max}) denoting 
maximal absolute value for $x \in \left[ x_0,x_1\right]$, so that the validity of the condition (C2$^\prime$) implies in the validity of (C2). Then, we set $s_{0}\!=\!0$ and define the adiabaticity coefficients 
\begin{subequations}
	\label{EqAdCoeffOS}
	\begin{align}
	\Xi_{\alpha \beta}^{(1)}(s) &\equiv \left \vert \frac{\tilde{F}_{\alpha\beta}(s)  e^{\tau\int_{0}^{s} \Gcal_{\alpha\beta}(s^{\prime})ds^{\prime}}}{\tau\Gcal_{\alpha\beta}(s)} \right \vert \text{ , } \label{EqAdCoeffOS1} \\
	\Xi_{\alpha \beta}^{(2)}(s) &\equiv \left \vert \frac{1}{\tau}\frac{d}{ds} \left[ \frac{\tilde{F}_{\alpha\beta}(s)}{\Gcal_{\alpha\beta}(s)} \right]e^{\tau\int_{0}^{s} \Gcal_{\alpha\beta}(s^{\prime}) ds^{\prime}}
	\right \vert \text{ , } \label{EqAdCoeffOS2}
	\end{align}
\end{subequations}
so that the conditions (C1) and (C2) can be expressed as
\begin{equation}
\Xi_{\alpha \beta} = \max \left\{ \max_{s\in[0,1]} \Xi_{\alpha \beta}^{(1)}(s) , \max_{s\in[0,1]} \Xi_{\alpha \beta}^{(2)}(s) \right\}\ll 1 \text{ . } \label{EqAdCondOpenSystem-2}
\end{equation}
Notice that Eq.~(\ref{EqAdCondOpenSystem-2}) can be viewed as a generalization of the condition established in Ref.~\cite{Tong:07} for open quantum systems.
The nature of the function $\Gcal_{\alpha\beta}(t)$ needs to be addressed in details. In fact, since $\Gcal_{\alpha\beta}(t) \in \Cmath$, 
the argument in the exponential of the Eqs.~\eqref{EqAdCondOpenSystem} could admit both real and imaginary parts. On the one hand, the imaginary part of 
$\Gcal_{\alpha\beta}(t)$ can be neglected due to the absolute value in the adiabaticity coefficients. 
On the other hand, the real part of $\Gcal_{\alpha\beta}(t)$ may lead to the divergence of the exponencial in Eq.~\eqref{EqAdCondOpenSystem} for long evolution times. 
Therefore, adiabaticity is not generally achieved in the regime $\tau \rightarrow \infty$, but it may be possible to find an evolution time range for $\tau$ so that the 
adiabatic approximation can be successfully implemented~\cite{Sarandy:05-1,Sarandy:05-2,Sarandy:04}. 

For the general case of multidimensional Jordan blocks in Eq.~\eqref{JDiago}, one needs to start from the coupled set of equations given in Eq.~\eqref{Eqrdot}. 
Without loss of generality, let us define the most general parameter $p_{\beta}^{k}(t)$ through 
\begin{eqnarray}
\dot{r}_{\beta}^{k}(t) = p_{\beta}^{k}(t) e^{\int_{t_{0}}^{t} \lambda_{\beta}(\xi)d\xi} \text{ , }
\end{eqnarray}
such that Eq.~\eqref{Eqrdot} becomes
\begin{eqnarray}
&&\dot{p}_{\beta}^{k}(t) = - p_{\beta}^{k}(t) \dinterpro{\Ecal_{\beta}^{k}(t)}{\dot{\Dcal}_{\beta}^{k}(t)}  
- \sum _{n_{\beta} \neq k} p_{\beta}^{n_{\beta}}(t) \dinterpro{\Ecal_{\beta}^{k}(t)}{\dot{\Dcal}_{\beta}^{k}(t)} \hspace{0.8cm} \nonumber \\ 
&& + p_{\beta}^{k+1}(t) - \sum_{\alpha\neq \beta} \sum _{n_{\alpha}} p_{\alpha}^{n_{\alpha}}(t) e^{\int_{t_{0}}^{t} \left[\lambda_{\alpha}(\xi) - \lambda_{\beta}(\xi)\right] d\xi}  \dinterpro{\Ecal_{\beta}^{k}(t)}{\dot{\Dcal}_{\alpha}^{n_{\alpha}}(t)} \text{ .} \label{Eqpdot}
\end{eqnarray}
Eq.~(\ref{Eqpdot}) describes the dynamics for the $k$-th vector in the $\beta$-th Jordan block. 
As previously mentioned, the last sum term in the right-hand-side of Eq.~(\ref{Eqpdot}) is the `diabatic' 
contribution to the dynamics, which couples distinct Jordan blocks. 
Therefore, by imposing adiabaticity, Eq.~(\ref{Eqpdot}) reduces to
\begin{align}
\dot{p}_{\beta}^{k}(t) & = - p_{\beta}^{k}(t) \dinterpro{\Ecal_{\beta}^{k}(t)}{\dot{\Dcal}_{\beta}^{k}(t)} + p_{\beta}^{k+1}(t) 
\nonumber \\ 
&- \sum _{n_{\beta} \neq k} p_{\beta}^{n_{\beta}}(t) \dinterpro{\Ecal_{\beta}^{k}(t)}{\dot{\Dcal}_{\beta}^{n_{\beta}}(t)} \text{ , } \label{Dotpk}
\end{align}
By following the same procedure as before, we can show that a sufficient condition for the adiabatic approximation in the case of multidimensional Jordan blocks is provided by~(See Appendix~\ref{ApConditions})
\begin{subequations}\label{EqAdCondOpenSystemGenBlock}
	\begin{align}
	(\overline{\text{C1}}) ~~ & \left \vert \frac{\tilde{F}^{k}_{\alpha\beta}(s)  e^{\tau\int_{s_{0}}^{s} \Gcal_{\alpha\beta}(s^{\prime})ds^{\prime}}}{\tau\Gcal_{\alpha\beta}(s)} \right \vert \ll 1 \text{ , } \label{EqACOpenGenBlockC1} \\
	(\overline{\text{C2}}) ~~ & \left \vert \frac{1}{\tau}\int_{s_{0}}^{s} \frac{d}{ds^{\prime}} \left[ \frac{\tilde{F}^{k}_{\alpha\beta}(s^{\prime})}{\Gcal_{\alpha\beta}(s^{\prime})} \right]e^{\tau\int_{s_{0}}^{s^{\prime}} \Gcal_{\alpha\beta}(s^{\prime\prime}) ds^{\prime\prime}} ds^{\prime} \right \vert \ll 1 \text{ , } \label{EqACOpenGenBlockC2}
	\end{align}
\end{subequations}
where $\Gcal(s) = \lambda_\alpha(s) - \lambda_\beta(s) \ne 0$ and the term $\tilde{F}^{k}_{\alpha\beta}(s)$ generalizes Eq.~\eqref{EqFalphaBeta} as
	\begin{equation}
	\tilde{F}^{k}_{\alpha\beta}(s) = \sum _{n_{\alpha} = 1}^{N_{\alpha}} e^{-\int_{s_{0}}^{s} \dinterpro{\Ecal_{\beta}^{k}(s^{\prime})}{d_{s^{\prime}}\Dcal_{\beta}^{k}(s^{\prime})} ds^{\prime} } \dinterpro{\Ecal_{\beta}^{k}(s)}{d_{s}\Dcal_{\alpha}^{n_{\alpha}}(s)} .
	\end{equation}
As in the case of one-dimensional blocks, conditions above are required to hold for every $s_0$ such that $s_0 \leq s \leq 1$. Similarly as in Eq.~(\ref{EqACOpenC2Max}), 
we can also rewrite condition $(\overline{\text{C2}})$ as
\begin{equation}
(\overline{\text{C2}^{\prime}}) \quad \quad \left \vert \frac{1}{\tau} \frac{d}{ds^{\prime}} \left[ \frac{\tilde{F}^{k}_{\alpha\beta}(s^{\prime})}{\Gcal_{\alpha\beta}(s^{\prime})} \right]e^{\tau\int_{s_{0}}^{s^{\prime}} \Gcal_{\alpha\beta}(s^{\prime\prime}) ds^{\prime\prime}}
\right \vert_M  \ll 1 \text{ . } \label{EqACOpenGenC2Max}
\end{equation}
We emphasize that the conditions presented here are \textit{sufficient} for ensuring the adiabatic approximation in open system, but they are not \textit{necessary} in general. 
In addition, we remark that these conditions also predict a possible adiabaticity breaking at finite time, as first discussed in Refs.~\cite{Sarandy:05-1,Sarandy:05-2}. 
However, as it has been shown in Ref.~\cite{Hu:19-a}, when additional conditions on the initial state of the system are satisfied, such behavior is suppressed 
and the open system adiabatic approximation can be achieved for arbitrary slowly varying dynamics.

\subsection{The adiabatic evolution superoperator in open systems} \label{SecU}

In this section, we derive a non-unitary evolution superoperator for the adiabatic open quantum dynamics driven by invertible dynamical maps. 
To this end, let us start with the case of Lindblad superoperators that admit one-dimensional Jordan block decomposition. 
Then, consider an initial state $\dket{\rho(t_{0})}$ written in the basis $\{\dket{\Dcal_{\alpha}(t_{0})}\}$ as
\begin{align}
\dket{\rho(t_{0})} = \sum\nolimits_{\alpha = 0}^{N-1} r_{\alpha}(t_{0}) \dket{\Dcal_{\alpha}(t_{0})} \text{ . }
\end{align}
If the system evolves through an adiabatic path in open system, we can use the Eq.~\eqref{EqAdEvol1D} to write the non-unitary evolution superoperator as
\begin{align}
\Ucal_{\text{ad}}^{\text{1D}}(t,t_{0}) = \sum_{\alpha = 0}^{N-1} e^{\int_{t_{0}}^{t} \Lambda_{\alpha}(\xi)d\xi}\dket{\Dcal_{\alpha}(t)}\dbra{\Ecal_{\alpha}(t_{0})} \text{ . } \label{OpEvo1D}
\end{align}
It is straightforward to show that Eq.~(\ref{OpEvo1D}) allows us to write $\dket{\rho^{1\text{D}}_{\text{ad}}(t)}\!=\!\Ucal_{\text{ad}}^{\text{1D}}(t,t_{0})\dket{\rho(t_{0})}$. 
The non-unitarity of the evolution naturally leads to a non-unitary $\Ucal_{\text{ad}}^{\text{1D}}(t,t_{0})$. 
However, it is possible to find an inverse superoperator $[\Ucal_{\text{ad}}^{\text{1D}}(t,t_{0})]^{-1}$ such that 
\begin{equation}
[\Ucal_{\text{ad}}^{\text{1D}}(t,t_{0})]^{-1}\Ucal_{\text{ad}}^{\text{1D}}(t,t_{0})=\Ucal_{\text{ad}}^{\text{1D}}(t,t_{0})[\Ucal_{\text{ad}}^{\text{1D}}(t,t_{0})]^{-1} = \1 \text{ .} \label{InvOpEvo1D}
\end{equation}
The inverse superoperator can be explicitly built upon the bi-orthonormalization condition obeyed by the basis vectors 
$\{\dket{\Dcal_{\alpha}(t)}\}$ and $\{\dbra{\Ecal_{\alpha}(t)}\}$, reading
\begin{eqnarray}
[\Ucal_{\text{ad}}^{\text{1D}}(t,t_{0})]^{-1} = \sum_{\alpha = 0}^{N-1} e^{-\int_{t_{0}}^{t} \Lambda_{\alpha}(\xi)d\xi}\dket{\Dcal_{\alpha}(t_{0})}\dbra{\Ecal_{\alpha}(t)} \text{ . }
\end{eqnarray}
From Eqs.~\eqref{OpEvo1D},~\eqref{InvOpEvo1D}, and~\eqref{EqEqEigenOneLa}, it follows that
\begin{equation}
[\Ucal_{\text{ad}}^{\text{1D}}(t,t_{0})]^{-1}\Lmath(t)\Ucal_{\text{ad}}^{\text{1D}}(t,t_{0}) = \sum_{\alpha = 0}^{N-1} \lambda_{\alpha}(t)\dket{\Dcal_{\alpha}(t_{0})}\dbra{\Ecal_{\alpha}(t_{0})} \text{ , }
\label{UOSdiag}
\end{equation}
Then, the superoperator $\Ucal_{\text{ad}}^{\text{1D}}(t,t_{0})$ diagonalizes the Lindbladian in the time-independent basis $\{\dket{\Dcal_{\alpha}(t_{0})},\dbra{\Ecal_{\alpha}(t_{0})}\}$. 
In the context of closed systems, this kind of result has shown useful applications in shortcuts to adiabaticity, such as the definition of multiple Schr\"odinger pictures~\cite{Ibanez:12} 
(or adiabatic iteration~\cite{Garrido:64,Berry:87}). As we shall see, the result can be generalized to the case of multidimensional Jordan blocks.
To extend Eq.~(\ref{UOSdiag}) for multidimensional Jordan blocks, we proceed by rewriting Eq.~\eqref{Dotpk} as
\begin{eqnarray}
\dot{\vec{p}}_{\beta}(t) = \left[ \tilde{\1}_{\text{u-shift}} - \Gcal_{\beta}(t) \right] \vec{p}_{\beta}(t) \text{ . } \label{Eqpvec}
\end{eqnarray}
Here $\Gcal_{\beta}(t)$ is a $(N_{\beta}\times N_{\beta})$-dimensional matrix whose elements are $\Gcal_{\beta}^{kn}(t) = \dinterpro{\Ecal_{\beta}^{k}(t)}{\dot{\Dcal}_{\beta}^{n}(t)}$, 
$\vec{p}_{\beta}(t)$ is a vector with $N_{\beta}$ components $p_{\beta}^{k}(t)$, and $\tilde{\1}_{\text{u-shift}}$ is an upper shift matrix 
\begin{eqnarray}
\tilde{\1}_{\text{u-shift}} = \begin{bmatrix}
0 & 1 & 0 & \cdots & 0 \\
0 & 0 & 1 & \cdots & 0 \\
\vdots & \ddots & \ddots & \ddots & \vdots \\
0 & \cdots & \cdots & 0 & 1 \\
0 & \cdots & \cdots & 0 & 0
\end{bmatrix} \text{ . }
\end{eqnarray}
Thus, it follows that a decoupled evolution within a single Jordan block is not necessarily obtained even if the off-diagonal elements of $\Gcal_{\beta}(t)$ can be neglected. 
A convenient way to express the adiabatic evolution superoperator $\Ucal (t,t_{0})$ is by using the definition of evolution superoperators for individual blocks $\Ucal_{\beta} (t,t_{0})$, reading
\begin{equation}
\Ucal_{\beta} (t,t_{0}) = e^{\int_{t_{0}}^{t} \lambda_{\beta}(\xi)d\xi} \sum _{n_{\beta} = 1}^{N_{\beta}} \sum _{m_{\beta} = 1}^{N_{\beta}} v_{n_{\beta}m_{\beta}}(t)\dket{\Dcal_{\beta}^{n_{\beta}}(t)}\dbra{\Ecal_{\beta}^{m_{\beta}}(t_{0})} \label{EqUalpha}\text{ , }
\end{equation}
where the elements $v_{n_{\beta}m_{\beta}}(t)$ account for inner transitions within a single Jordan block. 
The functions $v_{n_{\beta}m_{\beta}}(t)$ are determined by Eq.~\eqref{Eqpvec}, therefore they depend on the elements of the matrix $\Gcal_{\beta}(t)$. 
From this definition, the complete evolution superoperator $\Ucal (t,t_{0})$ is given by
\begin{eqnarray}
\Ucal_{\text{ad}} (t,t_{0}) = \sum_{\alpha = 0}^{N-1} \Ucal_{\alpha} (t,t_{0}) \text{ . } \label{EqUAdOS}
\end{eqnarray}
Notice that, as expected, $\Ucal_{\text{ad}} (t,t_{0})$ does not admit transitions between two vectors from different blocks. 
It is important to mention the existence of an inverse superoperator $\Ucal_{\text{ad}}^{-1} (t,t_{0})$ 
such that $\Ucal_{\text{ad}}  (t,t_{0})\Ucal_{\text{ad}} ^{-1} (t,t_{0})\!=\!\1$. The superoperator $\Ucal_{\text{ad}} ^{-1} (t,t_{0})$ can be 
explicitly provided by
\begin{eqnarray}
\Ucal_{\text{ad}}^{-1} (t,t_{0}) = \sum_{\alpha = 0}^{N-1} \Ucal_{\alpha}^{-1} (t,t_{0}) \text{ , } \label{UOSinv}
\end{eqnarray}
where
\begin{equation}
\Ucal_{\alpha}^{-1} (t,t_{0}) = e^{-\int_{t_{0}}^{t} \lambda_{\alpha}(\xi)d\xi} \sum _{n_{\alpha} = 1}^{N_{\alpha}} \sum _{m_{\alpha} = 1}^{N_{\alpha}} \tilde{v}_{n_{\alpha}m_{\alpha}}(t)\dket{\Dcal_{\alpha}^{n_{\alpha}}(t_{0})}\dbra{\Ecal_{\alpha}^{m_{\alpha}}(t)} \text{ , } \label{UinverGenBlock}
\end{equation}
with the coefficients $\tilde{v}_{n_{\alpha}m_{\alpha}}(t)$ and $v_{n_{\beta}m_{\beta}}(t)$ obeying
\begin{eqnarray}
\sum _{j_{\nu} = 1}^{N_{\nu}} 
v_{\ell_{\nu}j_{\nu}}(t)\tilde{v}_{j_{\nu}m_{\nu}}(t) &=& \delta_{\ell_{\nu}m_{\nu}} \text{ , } \label{EqUcoeff} \\
\sum _{j_{\nu} = 1}^{N_{\nu}} 
\tilde{v}_{\ell_{\nu}j_{\nu}}(t){v}_{j_{\nu}m_{\nu}}(t) &=& \delta_{\ell_{\nu}m_{\nu}} \text{ . } \label{EqUcoeffINV}
\end{eqnarray}
In addition, the operator $\Ucal_{\text{ad}} (t,t_{0})$ can be identified as the superoperator that ``block-diagonalizes'' $\Lmath(t)$, 
which is achieved by using the additional constraint
\begin{equation}
\sum _{n_{\nu} = 1}^{N_{\nu}} \tilde{v}_{g_{\nu}(n_{\nu}-1)} v_{n_{\nu}l_{\nu}} = \delta_{l_\nu\, (g_\nu+1)} , 
\label{Eqmu}
\end{equation}
with $\tilde{v}_{{g_{\nu}0}}\equiv 0$. By making use of the constraints over $\Ucal_{\alpha}(t,t_{0})$, we can then show that 
\begin{eqnarray}
\Lmath_{\text{J}}(t) = \Ucal_{\text{ad}}^{-1} (t,t_{0}) \Lmath(t) \Ucal_{\text{ad}} (t,t_{0}) \text{ . } \label{EqLDiagU}
\end{eqnarray}
A detailed proof of Eqs.~\eqref{EqUcoeff},~\eqref{EqUcoeffINV}, and \eqref{Eqmu} is provided in Appendix~\ref{ApUAd}, where we show that  
Eq.~\eqref{UinverGenBlock} implies that $\Lmath_{\text{J}}(t)$ is block diagonal in the time-independent vector bases 
$\{\dket{\Dcal_{\alpha}^{n_{\alpha}}(t_{0})}\}$ and $\{\dbra{\Ecal_{\alpha}^{n_{\alpha}}(t_{0})}\}$.

\section{Applications} \label{SecApplications}

\subsection{Adiabatic quantum thermodynamics}

As a first application, let us now show that the sufficient conditions for adiabaticity in open systems are compatible with quantum thermodynamics at equilibrium. 
Notice that the standard adiabatic theorem is not generally applicable to the dynamics of a quantum system evolving at equilibrium in contact with a thermal reservoir, since 
the original adiabatic theorem is derived for unitary evolution. This analysis can be rigorously implemented here, since we are dealing with an adiabatic approximation 
derived for open quantum systems. 

Consider a quantum system driven by a time-dependent Hamiltonian $H(t)$ and in permanent contact with a thermal bath at temperature $T$. 
The system is initially prepared in an equilibrium state $\rho(0) = \exp[-\beta H(0)]/Z(0)$, where $Z(0)=\tr{\exp[-\beta H(0)]}$ is the partition function and $\beta=1/(kT)$, with 
$k$ denoting the Boltzmann constant. Assuming that the system slowly evolves through an equilibrium trajectory, the system will continuously relax to the 
instantaneous steady state $\rho_{\text{ss}}$ of the corresponding dynamical generator $\Lcal[\bullet]$ ($\Lcal[\rho_{\text{ss}}]\!=\!0$) .
By assuming evolution at thermal equilibrium, no  
entropy production is expected to occur, which means that entropy variation $dS$ is simply $dS = \beta dQ$, where $dQ$ the heat exchanged in the thermal process. 
Let us now show that this is indeed the case for a general adiabatic evolution of an open system describing the dynamics at thermal equilibrium. 
Let $\dket{\rho (0)} \!=\!\sum \nolimits _{i,k_{i}} c^{(k_{i})}_{i}\dket{\Dcal^{(k_{i})}_{i}(0)}$ be the initial state of the system. 
By considering a general adiabatic evolution, which is ensured by Eqs.~(\ref{EqACOpenGenBlockC1})~and~(\ref{EqACOpenGenBlockC2}), we have
\begin{align}
\dket{\rho^{\text{ad}}(t)} = \sum \nolimits _{i,k_{i}} c^{(k_{i})}_{i} e^{\int_{0}^{t} \tilde{\lambda}_{i,k_{i}}(t^{\prime})dt^{\prime}}\dket{\Dcal^{(k_{i})}_{i}(t)} \label{AdEvolvedAp}
\end{align}
where $\tilde{\lambda}_{i,k_{i}}(t^{\prime})\!=\!\lambda_i(t) - \dinterpro{\Ecal_{i}^{k_{i}}(t)}{\dot{\Dcal}^{(k_{i})}_{i}(t)}$.
From Ref.~\cite{Hu:20-a}, heat $dQ^{\text{ad}}$ and entropy variation $dS^{\text{ad}}$ for open system adiabatic dynamics can be expressed as 
\begin{eqnarray}
dQ^{\text{ad}} &=& \frac{1}{D}\sum \nolimits _{i,k_{i}} c^{(k_{i})}_{i} e^{\int_{0}^{t} \tilde{\lambda}_{i,k_{i}}(t^{\prime})dt^{\prime}}\dinterpro{h(t)}{\Lmath (t)|\Dcal^{(k_{i})}_{i}(t)}dt \text{ , } \label{QAd1} \\
dS^{\text{ad}} &=& - \frac{1}{D}\sum \nolimits _{i,k_{i}} c^{(k_{i})}_{i} e^{\int_{0}^{t} \tilde{\lambda}_{i,k_{i}}(t^{\prime})dt^{\prime}}\dbra{\rho^{\text{ad}}_{\log}(t)}\Lmath (t) \dket{\Dcal^{(k_{i})}_{i}(t)} dt \text{ . } \,\,\,\,\,\,\,\,\, \label{SAd1}
\end{eqnarray} 
 In Eqs.~(\ref{QAd1}) and (\ref{SAd1}), the left vectors $\dbra{h(t)}$ and $\dbra{\rho^{\text{ad}}_{\log}(t)}$ have components provided by 
 $\tr{H(t) \sigma^j}$ and $\tr{\log[\rho^{\text{ad}}(t)] \sigma^j}$, respectively, where $\log x \equiv \ln x$ denotes the natural logarithm and 
 $\rho^{\text{ad}}(t)$ is the density operator in the adiabatic regime. The eigenvalue equation for 
 $H(t)$ is written as $H(t)|E_n(t)\rangle=E_n(t) |E_n(t)\rangle$. 
 
From Eqs.~(\ref{QAd1}) and (\ref{SAd1}), the absence of irreversible entropy production is then ensured by verifying the relation $\dbra{\rho^{\text{ad}}_{\log}(t)}\!=\!-\beta \dbra{h(t)}$, where 
$\rho^{\text{ad}}(t)$ is taken as the Gibbs state $\rho_{\text{eq}}(t)\!=\!\exp[-\beta H(t)]/Z$, with $Z\!=\!\tr{ \exp[-\beta H(t)]}$. The equality trivially holds for the left-vector component $j\!=\!0$. For $1\le j \le D^2-1$, we have
 \begin{eqnarray}
 \tr{\log[\rho^{\text{ad}}(t)] \sigma^j} &=& \sum_n \langle E_n | \log[\rho_{\text{eq}}(t)] \sigma^j |E_n\rangle, \nonumber \\
 &=& \sum_{n}  \log\{\exp[-\beta E_n(t)] /Z\}  \langle E_n | \sigma^j |E_n\rangle, \nonumber \\
 &=& -\beta \tr{H(t) \sigma^j}, 
 \end{eqnarray}
 where we have used $\tr{\sigma^j} = 0$ in the last equality. This result shows that the $j$-th component of the vector $\dbra{\rho^{\text{ad}}_{\log}(t)}$ can be written 
 in terms of the corresponding component of $\dbra{h(t)}$ as $\dbra{\rho^{\text{ad}}_{\log}(t)}_{j}\!=\!-\beta \dbra{h(t)}_{j}$. Hence, from Eqs.~(\ref{QAd1}) and~(\ref{SAd1}), 
 we have $dS^{\text{ad}}\!=\!\beta dQ^{\text{ad}}$. In conclusion, our sufficient conditions for adiabaticity allow for a simple verification that the adiabatic approximation in 
 open systems is compatible with quantum thermodynamics at equilibrium. 
 
\subsection{Deutsch algorithm under dephasing}

Let us consider now an application of the open system adiabatic dynamics in quantum computation. In this direction, let us analyze the adiabatic Deutsch algorithm under dephasing. 
The problem addressed in Deutsch's algorithm~\cite{Deutsch:85} is how to determinate whether a dichotomic real function $f: x\in\{0,1\} \rightarrow f(x)\in\{0,1\}$ is \textit{constant} (the output result $f(x)$ is the same regardless input value $x$) or \textit{balanced} (the output result $f(x)$ assumes different values according with the input value $x$). Thus, let us denote $\Ocal_{f}$ as the operator associated to an oracle, 
which computes $f$, given by~\cite{Sarandy:05-2}
\begin{eqnarray}
\Ocal_{f} = (-1)^{f(0)}\ket{0}\bra{0} + (-1)^{f(1)}\ket{1}\bra{1} \text{ . }
\end{eqnarray}
Thus, one can write the adiabatic Hamiltonian that implements the algorithm as
\begin{eqnarray}
H^{\text{DA}}(t) = U_{f}(t) H_{0} U_{f}^{\dagger}(t) \text{ , }
\end{eqnarray}
where $H_{0} = -\hbar \omega \sigma_{x}/2$ and $U_{f}(t) = \exp(i\frac{\pi}{2}\frac{t}{\tau}\Ocal_{f})$. At $t=0$ we have $H^{\text{DA}}(0) =H_{0}$, so that the initial input state is written as $\ket{\psi_{\text{inp}}} = \ket{+} = (1/\sqrt{2})(\ket{0}+\ket{1})$. By considering a closed system dynamics and by assuming the evolution is slow enough, the output state can be described with high probability by the ground state of $H^{\text{DA}}(t)$, reading
\begin{eqnarray}
\rho_{\text{cs}}^{\text{DA}}(t) = \frac{1}{2} \left[ \1 + g_{\text{c}}(t) \sigma_{x} - g_{\text{s}}(t) \sigma_{y} \right] \text{ , } \label{EqStateDAOp}
\end{eqnarray}
where $g_{\text{c}}(t) = \cos \left( \pi F t/2\tau \right)$, $g_{\text{s}}(t) = \sin \left( \pi F t/2\tau \right)$, and $F = 1-(-1)^{f(0)+f(1)}$. We observe that the subscript 
``cs" denotes that $\rho_{\text{cs}}^{\text{DA}}(t)$ is obtained from the adiabatic solution for closed systems.

Now, let us consider that the system is interacting with a surrounding environment. Let us assume that the system-environment interaction can be modeled by a 
Markovian phase damping channel, with rate $\gamma(t)$. The system evolution can then be described by
\begin{eqnarray}
\dot{\rho}(t) = - \frac{i}{\hbar} [H^{\text{DA}}(t),\rho(t)] + \gamma(t) \left[ \sigma_{z} \rho(t) \sigma_{z} - \rho(t) \right] \text{ . } \label{EqLindDephDA}
\end{eqnarray}
In order to study the adiabatic dynamics of the system, let us rewrite it in the superoperator formalism as
\begin{eqnarray}
\dket{\dot{\rho}(t)} = \Lmath^{\text{DA}}(t) \dket{\rho(t)} \text{ , }
\end{eqnarray}
where
\begin{align}
\Lmath^{\text{DA}}(t) = \begin{bmatrix}
0 & 0 & 0 & 0 \\ 0 & -2 \gamma & 0 & \omega g_{\text{s}}(t) \\ 0 & 0 & -2 \gamma & \omega g_{\text{c}}(t) \\ 0 & - \omega g_{\text{s}}(t) & - \omega g_{\text{c}}(t) & 0
\end{bmatrix} \text{ . }
\end{align}
The right eigenvectors of $\Lmath^{\text{DA}}(t)$ are (the superscript ``t" denotes transpose)
\begin{subequations}
	\label{EqDARightEigenVec}
\begin{align}
\dket{\Dcal^{\text{DA}}_{0}(t)} &= \begin{bmatrix} \text{ } 1 & 0 & 0 & 0\text{ } \end{bmatrix}^{\text{t}} \text{ , } \\
\dket{\Dcal^{\text{DA}}_{1}(t)} &= \begin{bmatrix} \text{ } 0 & -g_{\text{c}}(t) & g_{\text{s}}(t) & 0\text{ } \end{bmatrix}^{\text{t}} \text{ , } \\
\dket{\Dcal^{\text{DA}}_{2}(t)} &= \begin{bmatrix} \text{ } 0 & \Delta_{+}(t) g_{\text{s}}(t) & \Delta_{+}(t) g_{\text{c}}(t) & 1\text{ } \end{bmatrix}^{\text{t}} \text{ , } \\
\dket{\Dcal^{\text{DA}}_{3}(t)} &= \begin{bmatrix} \text{ } 0 & \Delta^{-1}_{+}(t) g_{\text{s}}(t) & \Delta^{-1}_{+}(t) g_{\text{c}}(t) & 1\text{ } \end{bmatrix}^{\text{t}} \text{ , }
\end{align}
\end{subequations}
while left eigenvectors are
\begin{subequations}
	\label{EqDALeftEigenVec}
	\begin{align}
	\dbra{\Ecal^{\text{DA}}_{0}(t)} &= \begin{bmatrix} \frac{}{} 1 & 0 & 0 & 0\text{ } \end{bmatrix} \text{ , } \\
	\dbra{\Ecal^{\text{DA}}_{1}(t)} &= \begin{bmatrix} \text{ } 0 & -g_{\text{c}}(t) & g_{\text{s}}(t) & 0\text{ } \end{bmatrix} \text{ , } \\
	\dbra{\Ecal^{\text{DA}}_{2}(t)} &= \frac{1}{2} \begin{bmatrix} \text{ } 0 & \omega g_{\text{s}}(t) & \omega g_{\text{c}}(t) & - \frac{\Delta_{-}(t)}{\sqrt{\gamma^2(t)-\omega^2}}\text{ } \end{bmatrix} \text{ , } \\
	\dbra{\Ecal^{\text{DA}}_{3}(t)} &= \frac{1}{2}\begin{bmatrix}
	\text{ } 0 & - \omega g_{\text{s}}(t) & -\omega g_{\text{c}}(t) & \frac{\Delta_{+}(t)}{\sqrt{\gamma^2(t)-\omega^2}}\text{ }
	\end{bmatrix} \text{ , }
	\end{align}
\end{subequations}
with eigenvalues $\lambda_{0}(t) = 0$, $\lambda_{1}(t) = -2 \gamma(t) $, $\lambda_{2}(t) = - \Delta_{+}(t) $ and $\lambda_{3}(t) = -\Delta_{-}(t)$, where $\Delta_{\pm}(t) = \gamma(t) \pm \sqrt{\gamma^2(t) - 4\omega^2}$. 
The non-degenerate spectrum of $\Lmath^{\text{DA}}(t)$ shows that $\Lmath^{\text{DA}}(t)$ exhibits one-dimensional Jordan blocks, so that the adiabatic behavior of the system can be obtained from the adiabatic solution given in Eq.~\eqref{EqAdEvol1D}. We write the density matrix associated with the initial state as $\rho^{\text{DA}}(0) = \ket{\psi_{\text{inp}}}\bra{\psi_{\text{inp}}} = \ket{+}\bra{+} = (1/2)(\1+\sigma_{x})$. 
In the superoperator formalism, we can then show that the initial state can be written as a linear combination of the vectors $\dket{\Dcal^{\text{DA}}_{0}(0)}$ and $\dket{\Dcal^{\text{DA}}_{1}(0)}$ as
\begin{eqnarray}
\dket{\rho^{\text{DA}}(0)} &= \begin{bmatrix} \text{ } 1 & 1 & 0 & 0\text{ } \end{bmatrix}^{\text{t}} = \dket{\Dcal^{\text{DA}}_{0}(0)} - \dket{\Dcal^{\text{DA}}_{1}(0)} \text{ . } \label{EqIniStaDA}
\end{eqnarray}
Notice that the initial state necessarily requires superposition of two distinct Jordan blocks, with eigenvalues $\lambda_0(t)$ and $\lambda_1(t)$. 
From Eq.~(\ref{OpEvo1D}), the open system adiabatic evolution operator $\Ucal_{\text{DA}}(s)$ reads
\begin{align}
\Ucal_{\text{DA}}(t,0) = \sum_{\alpha = 0}^{3} e^{\int_{0}^{t} \Lambda_{\alpha}(\xi)d\xi}\dket{\Dcal^{\text{DA}}_{\alpha}(t)}\dbra{\Ecal^{\text{DA}}_{\alpha}(0)} . \label{EvolDA}
\end{align}
From Eq.~(\ref{EvolDA}), we can write the evolved state $\dket{\rho_{\text{ad}}^{\text{DA}}(t)}\!=\! \Ucal_{\text{DA}}(t,0)\dket{\rho_{\text{ad}}^{\text{DA}}(0)}$ as
\begin{align}
\dket{\rho_{\text{ad}}^{\text{DA}}(t)} = \dket{\Dcal^{\text{DA}}_{0}(t)} - e^{-2\int_{t_{0}}^{t} \gamma(\xi)d\xi}\dket{\Dcal^{\text{DA}}_{1}(t)} , \label{EqDAAdEvolution} 
\end{align}
where we used that $\Lambda_{1}(t)\!=\!\lambda_{1}(t)\!=\!-2 \gamma(t)$, since $\dket{\Dcal^{\text{DA}}_{1}(t)}$ is a real vector and satisfies $\dinterpro{\Ecal_{1}(t)}{\dot{\Dcal}_{1}(t)}=0$. By using that $\dbra{\Ecal^{\text{DA}}_{0}(t)} = \dket{\Dcal^{\text{DA}}_{0}(t)}^{\text{t}}$, we can write $\Xi_{0\alpha}(t) = 0$ $\forall \alpha \neq \beta$. Therefore, $\Xi_{0} = 0$ and 
$\dket{\Dcal^{\text{DA}}_{0}(t)}$ evolves independently of the other eigenvectors. Now, by rewriting Eq.~(\ref{EqDAAdEvolution}) in matrix notation, we get
\begin{align}
\dket{\rho_{\text{ad}}^{\text{DA}}(t)} &= \begin{bmatrix}
\text{ } 1 & e^{-2\int_{t_{0}}^{t} \gamma(\xi)d\xi}g_{\text{c}}(t) & - e^{-2\int_{t_{0}}^{t} \gamma(\xi)d\xi}g_{\text{s}}(t) & 0\text{ } 
\end{bmatrix}^{\text{t}} . 
\label{EqAdSolDA}
\end{align}
From Eq.~(\ref{EqAdSolDA}), we can determine the components of the coherence vector associated with density matrix $\rho^{\text{DA}}(t)$, yielding
\begin{align}
\rho_{\text{ad}}^{\text{DA}}(t) &= \frac{1}{2} \left[ \1 + e^{-2\int_{t_{0}}^{t} \gamma(\xi)d\xi}g_{\text{c}}(t) \sigma_{x} - e^{-2\int_{t_{0}}^{t} \gamma(\xi)d\xi}g_{\text{s}}(t)\sigma_{y} \right] . \label{EqDAAdSolOS}
\end{align}
Notice that, in the limit $\gamma(t) \rightarrow 0$, we recover the density matrix for the unitary dynamics shown in Eq.~\eqref{EqStateDAOp}, where the output state reads (at $t=\tau$)~\cite{Sarandy:05-2}
\begin{align}
\lim_{\gamma(t) \rightarrow 0} \, \rho_{\text{ad}}^{\text{DA}}(\tau) &= \rho^{\text{DA}}_{\text{cs}}(\tau) = \frac{1}{2} \left[ \1 + (-1)^{f(0)+f(1)} \sigma_{x} \right] , \label{EqOutStateDAOp}
\end{align}
with $\cos ( \pi F/2) = (-1)^{f(0)+f(1)}$ for $f(x)\in\{0,1\}$, since $F = 1-(-1)^{f(0)+f(1)}$. The above solution is the output for an optimal (non-decohering) situation. 
The experimental implementation of the adiabatic Deutsch algorithm under phase damping has been implemented via trapped ions~\cite{Hu:19-a}, where the adiabatic behavior is asymptotically observed for a long evolution time.

\begin{figure}
	\centering
	\includegraphics[scale=0.32]{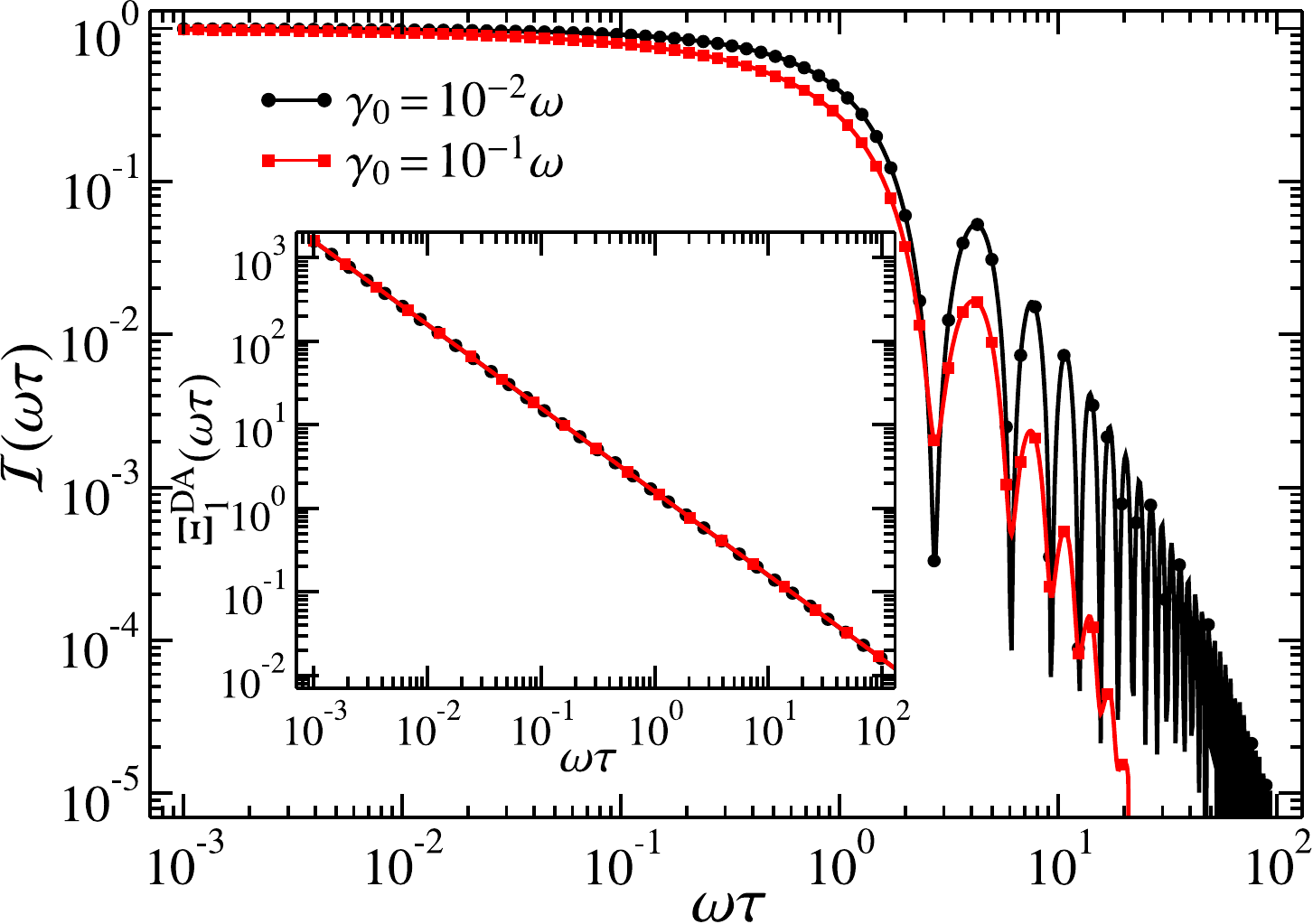}
	\caption{Infidelity $\Ical(\omega \tau)$ for achieving the open system adiabatic solution of the Deutsch algorithm for two different values of $\gamma_{0}$. Inset: The behavior of the the adiabaticity coefficient $\Xi_{1}^{\text{DA}}(\omega\tau)$. Here we consider the case where the function is balanced, {\it{i.e.}}, $F\!=\!2$.}
	\label{FigDATQD}
\end{figure}

In order to verify whether or not the sufficient conditions introduced here indicate the adiabatic behavior for the open system adiabatic Deutsch algorithm, we compute the fidelity~\cite{Nielsen:Book}
\begin{eqnarray}
\Fcal(\omega \tau) = \tr{\sqrt{\sqrt{\rho(\tau)}\rho^{\text{tar}}(\gamma_{0}\tau)\sqrt{\rho(\tau)}}} , \label{EqFidelOS}
\end{eqnarray}
where $\rho(\tau)$ is solution of the dynamics at $t=\tau$ and $\rho^{\text{tar}}$ is the target state. In our case, the target state is the adiabatic solution at $t=\tau$, obtained from Eq.~\eqref{EqDAAdSolOS} as
\begin{align}
\rho^{\text{DA}}(\gamma_{0}\tau) &= \frac{1}{2} \left[ \1 + e^{-2\gamma_{0}\tau}\cos \left( \frac{\pi F}{2}\right) \sigma_{x} - e^{-2\gamma_{0}\tau}\sin \left( \frac{\pi F}{2} \right)\sigma_{y} \right] . \label{EqDAAdSolOSEnd}
\end{align}
In Fig.~\ref{FigDATQD} we present the infidelity, $\Ical(\omega\tau)\!=\!1-\Fcal(\omega \tau)$, as a function of $\omega \tau$, since we set $\gamma_{0}$ as a multiple of $\omega$. 
Notice that the infidelity is asymptotically vanishing, decreasing faster for smaller rates $\gamma_0(t)$.  
However, it is important to highlight that an open system adiabatic high fidelity does not necessarily represent the solution for the Deutsch problem, because the state in Eq.~\eqref{EqDAAdSolOSEnd} 
is not the closed system final density operator~\cite{Hu:19-a}. Let us analyze now the behavior of the adiabaticity coefficient. We observe that 
$\Xi_{0n}^{\text{DA}}(\omega\tau)\!=\!\Xi_{n0}^{\text{DA}}(\omega\tau)\!=\!0$ for all $n$. Then, we focus on the nonvanishing coefficients for the Jordan block associated with 
$\dket{\Dcal^{\text{DA}}_{1}(t)}$. 
In this direction, we look at the quantities $\Xi_{1n}^{\text{DA}}(\omega\tau)$ and $\Xi_{n1}^{\text{DA}}(\omega\tau)$. 
Because we start the dynamics in a superposition of two different Jordan blocks and $\lambda_1\le\lambda_{n}$ $(n=2,3)$, the quantities $\Xi_{n1}^{\text{DA}}(\omega\tau)$ 
do not affect the dynamics since, from Eq.~(\ref{EqIniStaDA}), $\dket{\rho^{\text{DA}}(0)}$ does not depend on $\dket{\Dcal^{\text{DA}}_{2}(0)}$ and 
$\dket{\Dcal^{\text{DA}}_{3}(0)}$~\cite{Hu:19-a}. It is also possible to show that, for a constant dephasing rate $\gamma_{0}$, the contribution of the adiabatic condition (C2) 
trivially vanishes for the Deutsch algorithm. Indeed, we have $\Xi_{1 \beta}^{(2)}(s)\!=\!0$, since $\Gcal_{1\beta}(s)$ is constant, $\tilde{F}_{10}(s)\!=\!\tilde{F}_{11}(s)\!=\!0$, 
and $\tilde{F}_{1n}(s)\!=\!(-1)^n \omega \pi F / 4 \Delta$, where $\Delta^2\!=\!|\omega_{0}^{2} - \gamma_{0}^2|$ (for $n\!=\!2,3$). Hence, the derivative in Eq.~\eqref{EqAdCoeffOS2} vanishes. 
Concerning $\Xi_{1n}^{\text{DA}}(\omega\tau)$,
Eq.~(\ref{EqAdCoeffOS1}) then yields
\begin{align}
\Xi_{12}^{\text{DA}}(\omega\tau) = \Xi_{13}^{\text{DA}}(\omega\tau) = \max_{s\in[0,1]} \Xi_{1 \beta}^{(1)}(s) =  \frac{F\pi \omega e^{-\tau \gamma_{0}}}{4\tau \Delta |(i\Delta -\gamma_{0})| } .
\end{align}

We then define $\Xi_{1}^{\text{DA}}(\omega\tau)\!\equiv\! \max_{\beta\in\{0,2,3\}} [\Xi_{1\beta}^{\text{DA}}(\omega\tau)]$. 
The results for $\Xi_{1}^{\text{DA}}(\omega\tau)$ are shown in the inset of Fig.~\ref{FigDATQD}, allowing us to see the independent evolution of $\dket{\Dcal^{\text{DA}}_{1}(t)}$ as 
$\tau\!\rightarrow\!\infty$. 

\subsection{Landau-Zener under bit-phase-flip}

As another example of application of our adiabatic approach, let us consider the Landau-Zener Hamiltonian given by $
H_{\text{LZ}}(t) = \hbar\omega_{0} \sigma_{z} + \hbar\Delta(t) \sigma_{x}
$, where here we are considering a time-independent detuning frequency $\omega_{0}$. 
Let us assume that the system evolves under bit-phase flip decohering effect, whose the Lindblad equation reads
\begin{align}
\dot{\rho}(t) = - \frac{i}{\hbar} [H_{\text{LZ}}(t),\rho(t)] + \gamma(t) \left[ \sigma_{y} \rho(t) \sigma_{y} - \rho(t) \right] , \label{EqLindBitPPLZ}
\end{align}
where $\gamma(t)$ is the time-dependent bit phase flip decohering rate. Now, by writing the above equation in its superoperator form, we get (in the Pauli basis $\sigma_{i}=\{\1,\sigma_{x},\sigma_{y},\sigma_{z}\}$)
\begin{align}
\Lmath^{\text{LZ}}(t) = \begin{bmatrix}
0 & 0 & 0 & 0 \\ 
0 & -2 \gamma(t) & -\omega_{0} & 0 \\ 
0 & \omega_{0} & 0 & -\omega_{0}\tan\theta(t) \\ 
0 & 0 & \omega_{0}\tan\theta(t) & -2 \gamma(t)
\end{bmatrix} ,
\end{align}
where $\theta(t) = \arctan\left[\Delta(t)/\omega_{0}\right]$. The right eigenvectors are given by
\begin{subequations}
	\label{EqLZRightEigenVec}
	\begin{align}
	\dket{\Dcal^{\text{LZ}}_{0}(t)} &= \begin{bmatrix} \text{ } 1 & 0 & 0 & 0\text{ } \end{bmatrix}^{\text{t}} , \\
	\dket{\Dcal^{\text{LZ}}_{1}(t)} &= \begin{bmatrix} \text{ } 0 & \sin \theta(t) & 0 & \cos \theta(t) \text{ } \end{bmatrix}^{\text{t}} ,
	\\
	\dket{\Dcal^{\text{LZ}}_{2}(t)} &= \begin{bmatrix} \text{ } 0 & -\cos \theta(t) & \frac{\gamma(t)\cos \theta(t) - \kappa(t)}{\omega_{0}} & \sin \theta(t) \text{ } \end{bmatrix}^{\text{t}} ,
	\\
	\dket{\Dcal^{\text{LZ}}_{3}(t)} &= \begin{bmatrix} \text{ } 0 & -\cos \theta(t) & \frac{\gamma(t)\cos \theta(t) + \kappa(t)}{\omega_{0}} & \sin \theta(t) \text{ } \end{bmatrix}^{\text{t}} ,
	\end{align}
\end{subequations}
while the left eigenvectors are
\begin{subequations}
	\label{EqLZLeftEigenVec}
	\begin{align}
	\dbra{\Ecal^{\text{LZ}}_{0}(t)} &= \begin{bmatrix} \text{ } 1 & 0 & 0 & 0\text{ } \end{bmatrix} , \\
	\dbra{\Ecal^{\text{LZ}}_{1}(t)} &= \begin{bmatrix} \text{ } 0 & \sin \theta(t) & 0 & \cos \theta(t) \text{ } \end{bmatrix} ,
	\\
	\dbra{\Ecal^{\text{LZ}}_{2}(t)} &= \frac{1}{2} \begin{bmatrix} \text{ } 0 & -\cos \theta(t) \tilde{\kappa}_{+} & -\frac{\omega}{\kappa(t)} & \sin \theta(t) \tilde{\kappa}_{+} \text{ } \end{bmatrix} ,
	\\
	\dbra{\Ecal^{\text{LZ}}_{3}(t)} &= \frac{1}{2} \begin{bmatrix} \text{ } 0 & -\cos \theta(t) \tilde{\kappa}_{-} & \frac{\omega}{\kappa(t)} & \sin \theta(t) \tilde{\kappa}_{-} \text{ } \end{bmatrix} ,
	\end{align}
\end{subequations}
where we defined $\tilde{\kappa}_{\pm} =  1 \pm \cos \theta(t) \gamma(t)/\kappa(t)$, and $\kappa^2 (t) = \gamma^2(t)\cos^2 \theta(t)-\omega^2(t)$. 
The eigenvalues are $\lambda_{0}(t) = 0$, $\lambda_{1}(t) = -2 \gamma(t) $, and $\lambda_{n}(t) = -\gamma(t) -(-1)^{n}\kappa(t)$, where $n=\{2,3\}$. 

\begin{figure}[t!]
	\centering
	\includegraphics[scale=0.32]{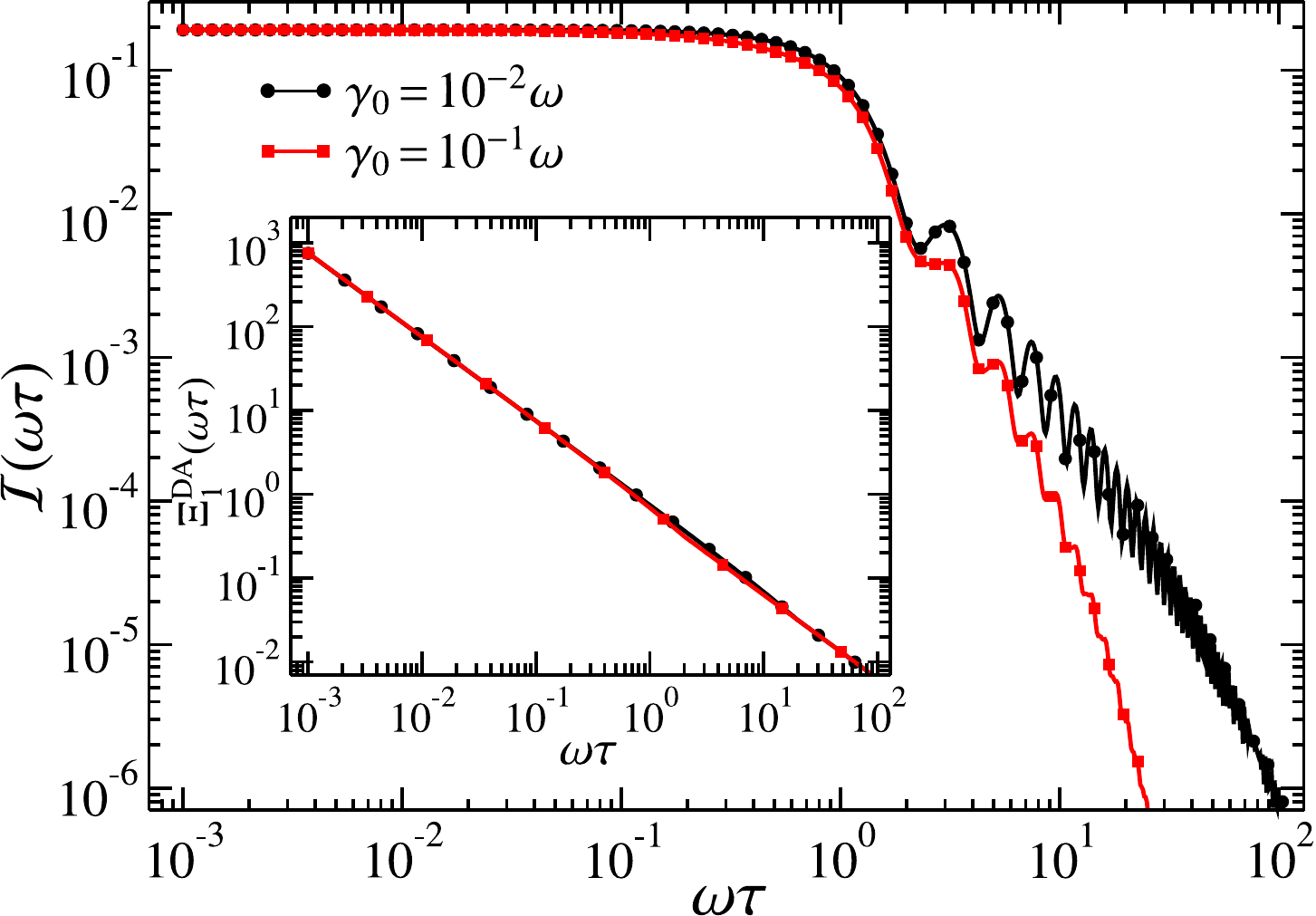}
	\caption{Infidelity $\Ical(\omega \tau)$ for achieving the open system adiabatic solution of the Landau-Zener dynamics under bit-phase-flip for different values of $\gamma_{0}$. 
	Inset: Adiabaticity coefficient $\Xi_{1}^{\text{LZ}}(\omega\tau)$. We have set $\theta\!=\!2\pi/5$.}
	\label{FigLZTQD}
\end{figure}

Now, by considering the case where $\Delta(0)=0$ and that the system is prepared in the ground state of $H(0)$, 
the initial state is given by $\rho^{\text{LZ}}(0) = \ket{1}\bra{1} = (1/2)(\1 - \sigma_{z})$ so that, in the superoperator formalism, we write
\begin{align}
\dket{\rho^{\text{LZ}}(0)} = \begin{bmatrix} \text{ } 1 & 0 & 0 & -1\text{ } \end{bmatrix}^{\text{t}} = \dket{\Dcal^{\text{LZ}}_{0}(0)} - \dket{\Dcal^{\text{LZ}}_{1}(0)} ,
\end{align}
where we already used that $\theta(0) = 0$ (since $\Delta(0)=0$) to write $\dket{\rho(0)}$ in terms of the eigenvectors of $\Lmath^{\text{LZ}}(0)$. 
Again, notice that the initial state necessarily requires superposition of two distinct Jordan blocks, with eigenvalues $\lambda_0(t)$ and $\lambda_1(t)$.
By assuming that the system undergoes adiabatic dynamics, the adiabatic evolution operator reads
\begin{align}
\Ucal_{\text{LZ}}(t,0) = \sum_{\alpha = 0}^{3} e^{\int_{0}^{t} \Lambda_{\alpha}(\xi)d\xi}\dket{\Dcal^{\text{LZ}}_{\alpha}(t)}\dbra{\Ecal^{\text{LZ}}_{\alpha}(0)} .
\end{align}	
The evolved state $\dket{\rho^{\text{LZ}}_{\text{ad}}(t,0)}\!=\!\Ucal_{\text{LZ}}(t)\dket{\rho^{\text{LZ}}(0)}$ is given by
\begin{align}
\dket{\rho^{\text{LZ}}_{\text{ad}}(t)} = \dket{\Dcal^{\text{LZ}}_{0}(t)} - e^{\int_{t_{0}}^{t} \Lambda_{1}(\xi) d\xi}\dket{\Dcal^{\text{LZ}}_{1}(0)} ,
\end{align}
where we now use $\dinterpro{\Ecal_{1}(t)}{\dot{\Dcal}_{1}(t)}\!=\!0$ to get $\Lambda_{1}(t)\!=\!-2\gamma(t)$. Thus, we write
\begin{align}
\dket{\rho^{\text{LZ}}_{\text{ad}}(t)} = \dket{\Dcal^{\text{LZ}}_{0}(t)} - e^{-\int_{t_{0}}^{t} 2 \gamma(\xi)d\xi}\dket{\Dcal^{\text{LZ}}_{1}(t)} , \label{lastLZ}
\end{align}
By rewriting Eq.~(\ref{lastLZ}) in the explicit vector notation, we obtain
\begin{align}
\dket{\rho^{\text{LZ}}_{\text{ad}}(t)} = \begin{bmatrix} \text{ } 1 & - e^{-\int_{t_{0}}^{t} 2 \gamma(\xi) d\xi}\sin \theta(t) & 0 & - e^{-\int_{t_{0}}^{t} 2 \gamma(\xi)d\xi} \cos \theta(t) \text{ } \end{bmatrix}^{\text{t}}
.
\end{align}
Therefore
\begin{align}
\rho^{\text{LZ}}_{\text{ad}}(t) = \frac{1}{2} \left[\1 - e^{-\int_{t_{0}}^{t} 2 \gamma(\xi) d\xi}\sin \theta(t) \sigma_{x} - e^{-\int_{t_{0}}^{t} 2 \gamma(\xi)d\xi} \cos \theta(t)\sigma_{z} \right]
.
\end{align}

\begin{figure}
	\centering
	\includegraphics[scale=0.27]{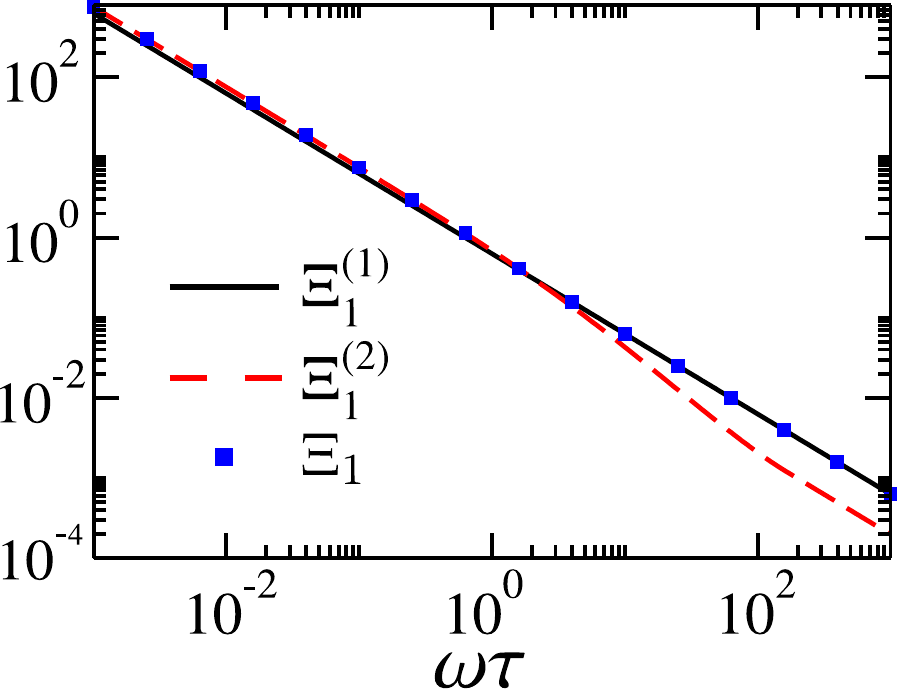}\label{logSc}~
	\includegraphics[scale=0.27]{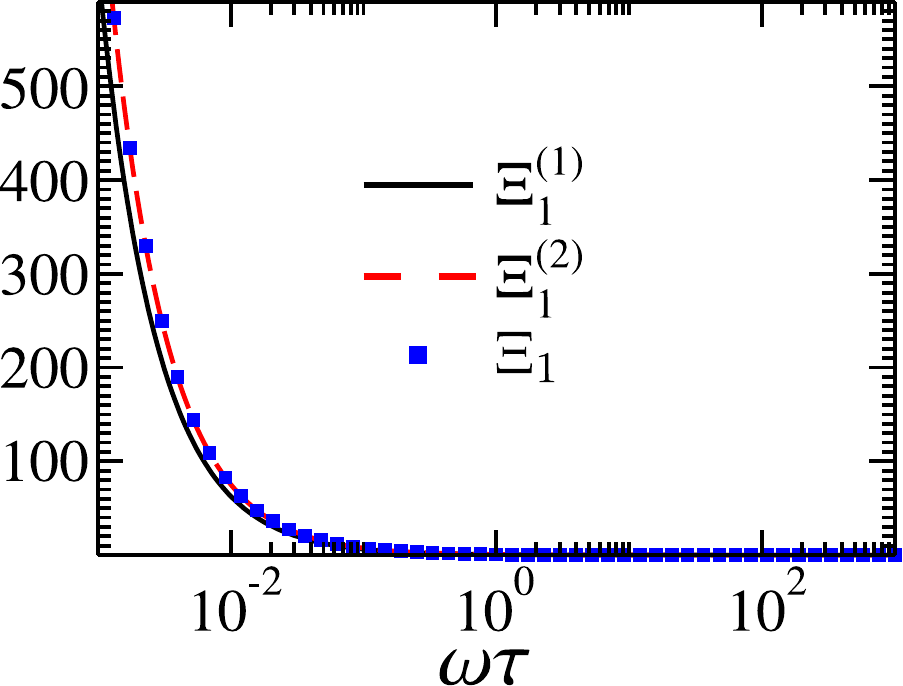}\label{lin}
	\caption{Quantities $\Xi_{1}^{(1)}$, $\Xi_{1}^{(2)}$, and $\Xi_{1}$ as a function of $\omega \tau$ for the Landau-Zener example in (left) log and (right) linear scale. 
	We have set $\gamma_{0} = 0.1\omega$ and $\theta\!=\!2\pi/5$.}\label{FigLZ-Conds}
\end{figure}

Similarly as before, we show in Fig.~\ref{FigLZTQD} the infidelity $\Ical(\omega \tau)$ for the Landau-Zener model. 
In agreement with the expected adiabatic behavior, the infidelity is asymptotically vanishing, decreasing faster for smaller rates $\gamma(t)$. 
Looking at the adiabatic coefficients, 
as in the previous example, it follows that the quantities $\Xi_{n1}^{\text{LZ}}(\omega\tau)$ do not affect the dynamics.
In addition, we define $\Xi_{1}^{\text{LZ}}(\omega\tau) \equiv \max_{n\in\{0,2,3\}} [\Xi_{1n}^{\text{LZ}}(\omega\tau)]$, 
where $\Xi_{1n}^{\text{LZ}}(\omega\tau)$ is given by Eq.~\eqref{EqAdCondOpenSystem-2} and takes into account the maximization over 
$\Xi_{1 n}^{(1)}(s)$ and $\Xi_{1n}^{(2)}(s)$, given by Eqs.~\eqref{EqAdCoeffOS1} 
and~\eqref{EqAdCoeffOS2}, respectively.
We numerically compute each coefficient $\Xi_{1n}^{\text{LZ}}(\omega\tau)$, exhibiting $\Xi_{1}^{\text{LZ}}(\omega\tau)$ in the inset of Fig.~\ref{FigLZTQD}. 
To show the nontrivial role played by conditions (C1) and  (C2), we explicitly plot in Fig.~\ref{FigLZ-Conds} the maximization over 
the individual adiabatic coefficients, which are denoted by $\Xi_{1}^{(k)}\!=\!\max_{s,\beta}[\Xi_{1 \beta}^{(k)}(s)]$ and $\Xi_{1}\!=\!\max_{k}[\Xi_{1}^{(k)}]$, with $k\!=\!1,2$. 
This plot shows that both conditions (C1) and (C2) are nontrivial for the Landau-Zener model and that 
the system approximately evolves through the adiabatic trajectory when the dynamics becomes sufficiently slow.

\section{Conclusions} \label{SecConc}

We derived new sufficient conditions for adiabaticity in open quantum systems. These conditions are simple yet general, allowing for the discussion of the adiabatic 
dynamics of arbitrary initial mixed states evolving in superpositions of Jordan eigenspaces driven by time local master equations.  In addition, we introduced a 
nonunitary adiabatic evolution superoperator $\Ucal_{\text{ad}} (t,t_{0})$, which has provided a convenient instrument to describe the open system dynamics.    
These results can be seen as an operational toolbox for investigating adiabaticity when environment effects become non-negligible. 

We illustrated the applications of our results in several distinct scenarios. First, we have shown that quantum systems evolving at thermal equilibrium can be 
consistently described in terms of the adiabatic approximation for open systems. Notice that the standard adiabatic theorem is derived explicitly for closed systems 
and cannot be applied in general to describe the dynamics of quantum systems interacting with an external environment. This problem is solved here by ensuring 
adiabaticity in open systems through the sufficient conditions derived in our work. Moreover, we also provided an illustration of our method in quantum control, 
evaluating the adiabatic behavior for the Hamiltonians of the Deutsch algorithm and the Landau-Zener model under decoherence. In both cases, state preparation 
has led to superpositions of basis vectors belonging to distinct Jordan subspaces, which required an adiabatic approximation beyond the single Jordan block picture. They have been 
analytically treated, with the asymptotic adiabatic trend explicitly shown through the behavior of the infidelity for the adiabatic state and through the vanishing of the 
adiabatic coefficients for long times. 

As a future perspective, 
we intend to consider corrections to the equilibrium state, which can be derived by 
considering a perturbative approach. For closed systems, the adiabatic approximation can be taken as 
the zeroth-order term in the adiabatic series expansion. Further 
corrections can then be obtained in terms of powers of the adiabatic parameter $1/T$, with $T$ denoting the total time of evolution 
(see, e.g., Refs.~\cite{Rigolin:08,Rigolin:10}). 
In the case of open systems under time-local evolution, we can similarly pursue an adiabatic power expansion, 
whose starting point can be the formal methods in Appendix~\ref{ApConditions}. 
Moreover, 
we also aim at applying the methods developed in this work to derive generalized shortcuts to adiabaticity in open quantum systems. In particular, 
the adiabatic evolution superoperator can be applied as a potential tool to provide non-transitional dynamics in terms of Jordan eigenspaces, similarly as considered by Vacanti {\it{et al.}} in 
Ref.~\cite{Vacanti:14}. This is expected to yield families of non-transitional trajectories in Hilbert-Schmidt space, allowing for the search of convenient setups in a 
reservoir engineering approach.

\begin{acknowledgments}
A. C. Santos acknowledges financial support through the research grant from the São Paulo Research Foundation (FAPESP) (Grant No 2019/22685-1).
M.S.S. acknowledges financial support from the Conselho Nacional de Desenvolvimento Científico e Tecnológico (CNPq-Brazil) (No. 303070/2016-1).  
This research is also supported in part by the Coordena\c{c}\~ao de Aperfei\c{c}oamento de Pessoal de N\'{\i}vel Superior - Brasil (CAPES) (Finance Code 001) and by the Brazilian
National Institute for Science and Technology of Quantum Information [CNPq INCT-IQ (465469/2014-0)]. \end{acknowledgments}

\appendix


\section{Superoperator formalism} \label{ApenSupForm}

Let us take a quantum system $S$ described by a $D_S$-dimensional Hilbert space $\Hcal_{\text{S}}$, which is coupled to a surrounding environment. 
We consider that the evolution of $S$ is  governed by a time-local master equation~\cite{Alicki:Book07}
\begin{eqnarray}
\dot{\rho}(t) = \Lcal_{t}[\rho(t)] \text{ , } \label{EqEqLind}
\end{eqnarray}
where $\Lcal_{t}[\bullet]$ is the generator of the dynamics and the subscript ``$t$'' makes explicit the possibility of time-dependence for $\Lcal_{t}[\bullet]$. 
By adopting the superoperator formalism, the density operator is taken as a $D_S^2$-dimensional vector $\dket{\rho(t)}$ in Hilbert-Schmidt space~\cite{Muynck:Book} 
(hence the double ket notation), with the dynamical generator being described by a $(D_S^2 \times D_S^2)$-dimensional {\it superoperator} $\Lmath(t)$, which acts on Hilbert-Schmidt space.   
In order to provide the components of $\dket{\rho(t)}$ and $\Lmath(t)$, 
we start by designing a matrix basis composed by the identity $ \1$ and a set of $(D_S^2 -1)$ operators $\{\sigma_{n}\}$ acting on $\Hcal_{\text{S}}$, 
with $\trs{\sigma_{n}} = 0$ and $\trs{\sigma_{n}\sigma_{m}^{\dagger}} = D_S\delta_{nm}$. 
In this basis, $\rho(t)$ can be written as
\begin{eqnarray}
\rho(t) = \frac{1}{D_S} \left[ \1 + \sum _{n=1}^{D_S^2 -1} \varrho_{n}(t) \sigma_{n} \right] \text{ , } \label{EqEqRhoCoherence}
\end{eqnarray}
with $\trs{\rho(t)} = 1$ and $\varrho_{n}(t) = \trs{\rho(t)\sigma_{n}^{\dagger}}$. 
For a two-level system, we can use $\{\1,\sigma_{n}\} = \{\1,\sigma_{x},\sigma_{y},\sigma_{z}\}$, with $\sigma_{n}$ denoting Pauli operators. 
Then, the density operator reads
\begin{eqnarray}
\rho(t) = \frac{1}{2} \left[ \1 + \varrho_{x}(t) \sigma_{x} + \varrho_{y}(t) \sigma_{y} + \varrho_{z}(t) \sigma_{z} \right] \text{ , }
\end{eqnarray}
with $\varrho_{n}(t)$ denoting the components of the three-dimensional \textit{coherence vector} $\vec{\varrho}(t)$. 
Returning to the $D^2_S$-dimensional case and using Eq.~\eqref{EqEqRhoCoherence} in Eq.~\eqref{EqEqLind}, we obtain
\begin{eqnarray}
\dot{\varrho}_{k} (t) = \frac{1}{D_S} \sum_{i=0}^{D_S^2-1} \trs{\sigma_{k}^{\dagger} \Lcal_t [ \sigma_{i} ]} \varrho_{i}(t) \text{ . } \label{Eqv1}
\end{eqnarray}
where $\Lcal [\bullet]$ is taken as a linear superoperator and we denote $\sigma_{0} = \1$. We now identify the coefficients $\trs{\sigma_{k}^{\dagger} \Lcal [ \sigma_{i} ]}$ 
in Eq.~(\ref{Eqv1}) as matrix elements at $k$-th row and $i$-th column of the $D_{S}^2 \times D_{S}^2$-dimensional superoperator $\Lmath(t)$. Then, we write 
\begin{eqnarray}
\dket{\dot{\rho}(t)} = \Lmath(t) \dket{\rho(t)} \text{ , } \label{ApEqSuperLindEq}
\end{eqnarray}
where $\dket{\rho(t)}$ is the $D_{S}^2$-dimensional \textit{coherence vector}, with components $\varrho_{n}(t) = \trs{\rho(t)\sigma_{n}^{\dagger}}$, $n=0,1,\cdots D_{S}^2-1$. 
In addition, the inner product between two density operators $\xi_{1}$ and $\xi_{2}$ can be provided in terms of their coherence vectors $\dket{\xi_{1}}$ and $\dket{\xi_{2}}$ 
as $\dinterpro{\xi_{1}}{\xi_{2}} = (1/D_S)\trs{\xi^{\dagger}_{1}\xi_{2}}$, where the dual coherence vector $\dbra{\xi_{1}}$ has components given by $\trs{\xi^{\dagger}_{1}\sigma_{n}}$.

\section{Adiabatic conditions on the total evolution time} \label{ApConditions}

Let us derive the conditions for adiabaticity shown in Eqs.~(\ref{EqACOpenC1})-(\ref{EqACOpenC2}) and Eqs.~(\ref{EqACOpenGenBlockC1})-(\ref{EqACOpenGenBlockC2}). 
First, let us start by considering the one-dimensional case by rewriting the Eq.~\eqref{EqpdotOne} as
\begin{widetext}
\begin{equation}
e^{\int_{t_{0}}^{t} \dinterpro{\Ecal_{\beta}(\xi)}{\dot{\Dcal}_{\beta}(\xi)} d\xi }\frac{d}{dt} \left[p_{\beta}(t) e^{- \int_{t_{0}}^{t} \dinterpro{\Ecal_{\beta}(\xi)}{\dot{\Dcal}_{\beta}(\xi)} d\xi }\right] = 
- \sum_{\alpha\neq \beta} p_{\alpha}(t) e^{\int_{t_{0}}^{t} \left[ \lambda_{\alpha}(\xi) - \lambda_{\beta}(\xi) \right] d\xi}\dinterpro{\Ecal_{\beta}(t)}{\dot{\Dcal}_{\alpha}(t)} \text{ , }
\end{equation}
so that one finds
\begin{equation}
p_{\beta}(t) e^{- \int_{t_{0}}^{t} \dinterpro{\Ecal_{\beta}(\xi)}{\dot{\Dcal}_{\beta}(\xi)} d\xi } - p_{\beta}(t_{0}) = 
- \sum_{\alpha\neq \beta} \int_{t_{0}}^{t} F_{\alpha\beta}(\xi)  e^{\int_{t_{0}}^{\xi} \left[ \lambda_{\alpha}(\xi^{\prime}) - \lambda_{\beta}(\xi^{\prime}) \right] d\xi^{\prime}} d\xi\text{ , }
\end{equation}
\end{widetext}
with
\begin{equation}
F_{\alpha\beta}(t) = e^{-\int_{t_{0}}^{t} \dinterpro{\Ecal_{\beta}(\xi)}{\dot{\Dcal}_{\beta}(\xi)} d\xi } p_{\alpha}(t) \dinterpro{\Ecal_{\beta}(t)}{\dot{\Dcal}_{\alpha}(t)} \text{ . }
\end{equation}
Therefore, the adiabatic approximation follows by requiring
\begin{equation}
G_{\alpha\beta}(t) = \left \vert \int_{t_{0}}^{t} F_{\alpha\beta}(\xi)  e^{\int_{t_{0}}^{\xi} \left[ \lambda_{\alpha}(\xi^{\prime}) - \lambda_{\beta}(\xi^{\prime}) \right] d\xi^{\prime}} d\xi \right \vert \ll 1 \text{ , } \label{Gabapp}
\end{equation}
for all $\alpha$, $\beta$ such that $\alpha \ne \beta$. Now, by using the normalized time $s = t/\tau$ in Eq.~(\ref{Gabapp}), we get
\begin{equation}
G_{\alpha\beta}(s) = \left \vert \int_{s_{0}}^{s} \tilde{F}_{\alpha\beta}(s^{\prime})  e^{\tau\int_{s_{0}}^{s^{\prime}} \left[ \lambda_{\alpha}(s^{\prime\prime}) - \lambda_{\beta}(s^{\prime\prime}) \right] ds^{\prime\prime}} ds^{\prime} \right \vert \text{ , }
\end{equation}
with 
\begin{equation}
\tilde{F}_{\alpha\beta}(s) = e^{-\int_{s_{0}}^{s} \dinterpro{\Ecal_{\beta}(s^{\prime})}{d_{s}\Dcal_{\beta}(s^{\prime})} ds^{\prime} } p_{\alpha}(s) \dinterpro{\Ecal_{\beta}(s)}{d_{s}\Dcal_{\alpha}(s)} \text{ , } \label{ApEqFalphaBeta}
\end{equation}
where we have adopted the notation $d_{s}f(s) \equiv df(s)/ds$.
By defining $\Gcal_{\alpha\beta}(s) = \lambda_{\alpha}(s) - \lambda_{\beta}(s)$ and assuming $\Gcal_{\alpha\beta}(s) \ne 0$, we can use 
\begin{align}
\frac{d}{ds} \left[ \frac{\tilde{F}_{\alpha\beta}(s)  e^{\tau\int_{s_{0}}^{s} \Gcal_{\alpha\beta}(s^{\prime})ds^{\prime}}}{\Gcal_{\alpha\beta}(s)} \right] &= \frac{d}{ds} \left[ \frac{\tilde{F}_{\alpha\beta}(s)}{\Gcal_{\alpha\beta}(s)} \right]e^{\tau\int_{s_{0}}^{s} \Gcal_{\alpha\beta}(s^{\prime}) ds^{\prime}} \nonumber \\&+ \tau \tilde{F}_{\alpha\beta}(s)  e^{\tau\int_{s_{0}}^{s} \Gcal_{\alpha\beta}(s^{\prime}) ds^{\prime}} \text{ , }
\end{align}
so that $G_{\alpha\beta}(s)$ can be rewritten as
\begin{align}
G_{\alpha\beta}(s) = &\left \vert \frac{1}{\tau} \int_{s_{0}}^{s} \frac{d}{ds^{\prime}} \left[ \frac{\tilde{F}_{\alpha\beta}(s^{\prime})  e^{\tau\int_{s_{0}}^{s^{\prime}} \Gcal_{\alpha\beta}(s^{\prime\prime})ds^{\prime\prime}}}{\Gcal_{\alpha\beta}(s^{\prime})} \right] 
\right. \nonumber \\
&\left.-\frac{d}{ds^{\prime}} \left[ \frac{\tilde{F}_{\alpha\beta}(s^{\prime})}{\Gcal_{\alpha\beta}(s^{\prime})} \right]e^{\tau\int_{s_{0}}^{s^{\prime}} \Gcal_{\alpha\beta}(s^{\prime\prime}) ds^{\prime\prime}}
ds^{\prime} \right \vert  \nonumber \\
= &\left \vert \frac{\tilde{F}_{\alpha\beta}(s)  e^{\tau\int_{s_{0}}^{s} \Gcal_{\alpha\beta}(s^{\prime})ds^{\prime}}}{\tau\Gcal_{\alpha\beta}(s)} - \frac{\tilde{F}_{\alpha\beta}(s_0) }{\tau\Gcal_{\alpha\beta}(s_0)} \right. \nonumber \\
&\left.
-\frac{1}{\tau}\int_{s_{0}}^{s} \frac{d}{ds^{\prime}} \left[ \frac{\tilde{F}_{\alpha\beta}(s^{\prime})}{\Gcal_{\alpha\beta}(s^{\prime})} \right]e^{\tau\int_{s_{0}}^{s^{\prime}} \Gcal_{\alpha\beta}(s^{\prime\prime}) ds^{\prime\prime}}
ds^{\prime} \right \vert \text{ . }
\end{align}
We then observe that we can achieve $G_{\alpha\beta}(s)\!\ll\!1$ by imposing the two conditions given by Eqs.~(\ref{EqACOpenC1})-(\ref{EqACOpenC2}). 
In a general case, where we have multidimensional Jordan blocks, one needs to start from Eq.~\eqref{Eqpdot}. Without loss of generality we write
\begin{equation}
r_{\beta}^{k}(t) = p_{\beta}^{k}(t) e^{\int_{t_{0}}^{t} \lambda_{\beta}(\xi)d\xi} \text{ , }
\end{equation}
where the argument $\int_{t_{0}}^{t} \lambda_{\beta}(\xi)d\xi$ can be associated with a dynamical phase (in analogy with closed systems) and Eq.~\eqref{Eqrdot} becomes
\begin{align}
\dot{p}_{\beta}^{k}(t) & = - p_{\beta}^{k}(t) \dinterpro{\Ecal_{\beta}^{k}(t)}{\dot{\Dcal}_{\alpha}^{k}(t)} 
- \sum _{n_{\beta} \neq k} p_{\beta}^{n_{\beta}}(t) \dinterpro{\Ecal_{\beta}^{k}(t)}{\dot{\Dcal}_{\beta}^{n_{\beta}}(t)} \nonumber \\ 
&+ p_{\beta}^{k+1}(t) - \sum_{\alpha\neq \beta} \sum _{n_{\alpha}} p_{\alpha}^{n_{\alpha}}(t) e^{\int_{t_{0}}^{t} \left[\lambda_{\alpha}(\xi) - \lambda_{\beta}(\xi) \right]d\xi}  \dinterpro{\Ecal_{\beta}^{k}(t)}{\dot{\Dcal}_{\alpha}^{n_{\alpha}}(t)} \text{ . } \label{ApEqpdot}
\end{align}
Therefore, by following the same procedure as before, we can show that the adiabatic dynamics is approximately achieved for
\begin{subequations}\label{ApEqAdCondOpenSystemGenBlock}
	\begin{align}
	(\overline{\text{C1}}) \quad \quad & \left \vert \frac{\tilde{F}^{k}_{\alpha\beta}(s)  e^{\tau\int_{s_{0}}^{s} \Gcal_{\alpha\beta}(s^{\prime})ds^{\prime}}}{\tau\Gcal_{\alpha\beta}(s)} \right \vert \ll 1 \text{ , } \label{ApEqACOpenGenBlockC1} \\
	(\overline{\text{C2}}) \quad \quad &\left \vert \frac{1}{\tau}\int_{s_{0}}^{s} \frac{d}{ds^{\prime}} \left[ \frac{\tilde{F}^{k}_{\alpha\beta}(s^{\prime})}{\Gcal_{\alpha\beta}(s^{\prime})} \right]e^{\tau\int_{s_{0}}^{s^{\prime}} \Gcal_{\alpha\beta}(s^{\prime\prime}) ds^{\prime\prime}}
	ds^{\prime} \right \vert \ll 1 \text{ , } \label{ApEqACOpenGenBlockC2}
	\end{align}
\end{subequations}
where $\Gcal_{\alpha\beta}(s) \ne 0$ and $\tilde{F}^{k}_{\alpha\beta}(s^{\prime})$ generalizes Eq.~\eqref{ApEqFalphaBeta} as
\begin{equation}
\tilde{F}^{k}_{\alpha\beta}(s) = \sum _{n_{\alpha} = 1}^{N_{\alpha}} p_{\alpha}^{n_{\alpha}}(s) e^{-\int_{s_{0}}^{s} \dinterpro{\Ecal_{\beta}^{k}(s^{\prime})}{d_{s^{\prime}}\Dcal_{\beta}^{k}(s^{\prime})} ds^{\prime} } \dinterpro{\Ecal_{\beta}^{k}(s)}{d_{s}\Dcal_{\alpha}^{n_{\alpha}}(s)} \text{ , }
\end{equation}
so that $\tilde{F}^{k}_{\alpha\beta}(s)$ reduces to $\tilde{F}_{\alpha\beta}(s)$ for one-dimensional Jordan blocks.

\section{The inverse of the adiabatic evolution superoperator} \label{ApUAd}

Let us derive the inverse of the evolution superoperator $\Ucal_{\text{ad}} (t,t_{0})$.  
To this end, we are required to find a superoperator $\Ucal_{\text{ad}}^{-1} (t,t_{0})$ such that $\Ucal_{\text{ad}} (t,t_{0})\Ucal_{\text{ad}}^{-1} (t,t_{0})=\1$. Then, let us define
\begin{eqnarray}
\Ucal_{\text{ad}}^{-1} (t,t_{0}) = \sum_{\alpha = 0}^{N-1} \Ucal_{\alpha}^{-1} (t,t_{0}) \text{ , }
\end{eqnarray}
where each contribution $\Ucal_{\alpha}^{-1} (t,t_{0})$ is taken as
\begin{align}
\Ucal_{\alpha}^{-1} (t,t_{0}) = e^{-\int_{t_{0}}^{t} \lambda_{\alpha}(\xi)d\xi} \sum _{n_{\alpha} = 1}^{N_{\alpha}} \sum _{m_{\alpha} = 1}^{N_{\alpha}} \tilde{v}_{n_{\alpha}m_{\alpha}}(t)\dket{\Dcal_{\alpha}^{m_{\alpha}}(t_{0})}\dbra{\Ecal_{\alpha}^{n_{\alpha}}(t)} \text{ , }
\end{align}
with parameters $\tilde{v}_{n_{\alpha}m_{\alpha}}(t)$ to be determined. This definition is convenient because we can write 
\begin{eqnarray}
\Ucal_{\beta} (t,t_{0})\Ucal_{\alpha}^{-1} (t,t_{0}) = \delta_{\alpha\beta} \Ucal_{\beta} (t,t_{0}) \Ucal_{\alpha}^{-1} (t,t_{0}) \text{ , }
\label{ApEqUortho}
\end{eqnarray}
where we use the bi-orthonormality relationship between right- and left-hand side quasi-eigenvectors. Now, we write
\begin{widetext}
\begin{align}
\Acal_{1} &= \Ucal_{\text{ad}} (t,t_{0})\Ucal_{\text{ad}}^{-1} (t,t_{0}) = \sum_{\alpha = 0}^{N-1} \sum_{\beta = 0}^{N-1} \Ucal_{\alpha} (t,t_{0}) \Ucal_{\beta}^{-1} (t,t_{0}) = \sum_{\alpha = 0}^{N-1} \Ucal_{\alpha} (t,t_{0}) \Ucal_{\alpha}^{-1} (t,t_{0}) \text{ , }
\end{align}
where we already used the Eq.~\eqref{ApEqUortho}. Thus
\begin{align}
\Acal_{1} &= \sum_{\alpha = 0}^{N-1} \left[ \left(\sum _{n_{\alpha} = 1}^{N_{\alpha}} \sum _{m_{\alpha} = 1}^{N_{\alpha}} v_{n_{\alpha}m_{\alpha}}(t)\dket{\Dcal_{\alpha}^{n_{\alpha}}(t)}\dbra{\Ecal_{\alpha}^{m_{\alpha}}(t_{0})}\right) 
\left(\sum _{j_{\alpha} = 1}^{N_{\alpha}} \sum _{k_{\alpha} = 1}^{N_{\alpha}} \tilde{v}_{j_{\alpha}k_{\alpha}}(t)\dket{\Dcal_{\alpha}^{j_{\alpha}}(t_{0})}\dbra{\Ecal_{\alpha}^{k_{\alpha}}(t)}\right)\right] \nonumber \\
&= \sum_{\alpha = 0}^{N-1} \left[ \sum _{n_{\alpha} = 1}^{N_{\alpha}} \sum _{m_{\alpha} = 1}^{N_{\alpha}}
\sum _{j_{\alpha} = 1}^{N_{\alpha}} \sum _{k_{\alpha} = 1}^{N_{\alpha}}
v_{n_{\alpha}m_{\alpha}}(t)\tilde{v}_{j_{\alpha}k_{\alpha}}(t)
\dket{\Dcal_{\alpha}^{n_{\alpha}}(t)}
\left( \dinterpro{\Ecal_{\alpha}^{m_{\alpha}}(t_{0})}{\Dcal_{\alpha}^{j_{\alpha}}(t_{0})}\right) \dbra{\Ecal_{\alpha}^{k_{\alpha}}(t)}\right] \nonumber \\
&= \sum_{\alpha = 0}^{N-1} \left[ \sum _{n_{\alpha} = 1}^{N_{\alpha}} \sum _{j_{\alpha} = 1}^{N_{\alpha}} \sum _{k_{\alpha} = 1}^{N_{\alpha}}
v_{n_{\alpha}j_{\alpha}}(t)\tilde{v}_{j_{\alpha}k_{\alpha}}(t)
\dket{\Dcal_{\alpha}^{n_{\alpha}}(t)}
\dbra{\Ecal_{\alpha}^{k_{\alpha}}(t)}\right] \text{ . }
\end{align}

Let us compute the matrix elements $\dbra{\Ecal_{\eta}^{\ell_{\eta}}(t)}\Acal_{1}\dket{\Dcal_{\nu}^{m_{\nu}}(t)}$. To simplify the notation, from now on, we will omit the time-dependence of the coefficients $v$ and $\tilde{v}$. Then
\begin{align}
\dbra{\Ecal_{\eta}^{\ell_{\eta}}(t)}\Acal_1 \dket{\Dcal_{\nu}^{m_{\nu}}(t)} &= \sum_{\alpha = 0}^{N-1} \left[ \sum _{n_{\alpha} = 1}^{N_{\alpha}} \sum _{j_{\alpha} = 1}^{N_{\alpha}} \sum _{k_{\alpha} = 1}^{N_{\alpha}}
v_{n_{\alpha}j_{\alpha}}\tilde{v}_{j_{\alpha}k_{\alpha}}
\dinterpro{\Ecal_{\eta}^{\ell_{\eta}}(t)}{\Dcal_{\alpha}^{n_{\alpha}}(t)}
\dinterpro{\Ecal_{\alpha}^{k_{\alpha}}(t)}{\Dcal_{\nu}^{m_{\nu}}(t)}\right] \nonumber \\
&= \sum_{\alpha = 0}^{N-1} \left[ \sum _{n_{\alpha} = 1}^{N_{\alpha}} \sum _{j_{\alpha} = 1}^{N_{\alpha}} \sum _{k_{\alpha} = 1}^{N_{\alpha}}
v_{n_{\alpha}j_{\alpha}}\tilde{v}_{j_{\alpha}k_{\alpha}}
\delta_{\ell_{\eta}n_{\alpha}} \delta_{k_{\alpha}m_{\nu}} \delta_{\eta\alpha}
\delta_{\alpha\nu}
\right] \nonumber \\
&= \sum_{\alpha = 0}^{N-1} \left[  \sum _{j_{\alpha} = 1}^{N_{\alpha}} 
v_{\ell_{\eta}j_{\alpha}}\tilde{v}_{j_{\alpha}m_{\nu}}
\delta_{\eta\alpha} \delta_{\alpha\nu}
\right] = \delta_{\eta\nu} \sum _{j_{\nu} = 1}^{N_{\nu}} 
v_{\ell_{\eta}j_{\nu}}\tilde{v}_{j_{\nu}m_{\nu}} \text{ . }
\end{align}
\end{widetext}
Therefore, to obtain $\Acal_{1} = \Ucal_{\text{ad}} (t,t_{0})\Ucal_{\text{ad}}^{-1} (t,t_{0}) = \1$, the coefficients are required to satisfy
\begin{eqnarray}
\sum _{j_{\nu} = 1}^{N_{\nu}} 
v_{\ell_{\nu}j_{\nu}}\tilde{v}_{j_{\nu}m_{\nu}} = \delta_{\ell_{\nu}m_{\nu}} \text{ . } \label{ApEqmu1}
\end{eqnarray}
Analogously, by requiring $\Ucal_{\text{ad}}^{-1} (t,t_{0}) \Ucal_{\text{ad}} (t,t_{0})=\1$, we obtain
\begin{eqnarray}
\sum _{j_{\nu} = 1}^{N_{\nu}} 
\tilde{v}_{\ell_{\nu}j_{\nu}}{v}_{j_{\nu}m_{\nu}} = \delta_{\ell_{\nu}m_{\nu}} \text{ . } \label{ApEqmuINV}
\end{eqnarray}
In addition, a further requirement for the operator $\Ucal_{\text{ad}} (t,t_{0})$ is the block-diagonalization of the Lindblad superoperator. 
Indeed, let us consider an operator $\Acal_{2}$ given by $\Acal_{2} = \Ucal_{\text{ad}}^{-1} (t,t_{0}) \Lmath(t) \Ucal_{\text{ad}} (t,t_{0})$, so that we have
\begin{align}
\Acal_{2} &= \Ucal_{\text{ad}}^{-1} (t,t_{0}) \Lmath(t) \Ucal_{\text{ad}} (t,t_{0}) \nonumber \\&= \sum_{\alpha = 0}^{N-1}\sum_{\beta = 0}^{N-1} 
\underbrace{\Ucal_{\alpha}^{-1} (t,t_{0}) \Lmath(t) \Ucal_{\beta} (t,t_{0})}_{\Acal_{2}^{\alpha\beta}} = \sum_{\alpha = 0}^{N-1}\sum_{\beta = 0}^{N-1} \Acal_{2}^{\alpha\beta} \text{ . }
\end{align}
Now observe we can write $\Acal_{2}^{\alpha\beta}$ as
\begin{widetext}
\begin{align}
\Acal_{2}^{\alpha\beta} &= \Ucal_{\alpha}^{-1} (t,t_{0}) \Lmath(t) \Ucal_{\beta} (t,t_{0}) 
\nonumber \\
&= 
e^{\int_{t_{0}}^{t} \left[ \lambda_{\beta}(\xi)-\lambda_{\alpha}(\xi) \right] d\xi  }
\sum _{j_{\alpha} = 1}^{N_{\alpha}} \sum _{k_{\alpha} = 1}^{N_{\alpha}}
\sum _{n_{\beta} = 1}^{N_{\beta}} \sum _{\ell_{\beta} = 1}^{N_{\beta}}
\tilde{v}_{j_{\alpha}k_{\alpha}} v_{n_{\beta}\ell_{\beta}}
\dbra{\Ecal_{\alpha}^{k_{\alpha}}(t)}\Lmath(t) \dket{\Dcal_{\beta}^{n_{\beta}}(t)}\dket{\Dcal_{\alpha}^{j_{\alpha}}(t_{0})}\dbra{\Ecal_{\beta}^{\ell_{\beta}}(t_{0})} 
\nonumber \\
&=
e^{\int_{t_{0}}^{t}  \left[  \lambda_{\beta}(\xi)-\lambda_{\alpha}(\xi) \right] d\xi }
\sum _{j_{\alpha} = 1}^{N_{\alpha}} \sum _{k_{\alpha} = 1}^{N_{\alpha}}
\sum _{n_{\beta} = 1}^{N_{\beta}} \sum _{\ell_{\beta} = 1}^{N_{\beta}}
\tilde{v}_{j_{\alpha}k_{\alpha}} v_{n_{\beta}\ell_{\beta}}\lambda_{\alpha}(t)
\dinterpro{\Ecal_{\alpha}^{k_{\alpha}}(t)}{\Dcal_{\beta}^{n_{\beta}}(t)}
\dket{\Dcal_{\alpha}^{j_{\alpha}}(t_{0})}\dbra{\Ecal_{\beta}^{\ell_{\beta}}(t_{0})}
\nonumber\\
&+
e^{\int_{t_{0}}^{t} \left[ \lambda_{\beta}(\xi)-\lambda_{\alpha}(\xi)\right] d\xi }
\sum _{j_{\alpha} = 1}^{N_{\alpha}} \sum _{k_{\alpha} = 1}^{N_{\alpha}}
\sum _{n_{\beta} = 1}^{N_{\beta}} \sum _{\ell_{\beta} = 1}^{N_{\beta}}
\tilde{v}_{j_{\alpha}k_{\alpha}} v_{n_{\beta}\ell_{\beta}}
\dinterpro{\Ecal_{\alpha}^{k_{\alpha}}(t)}{\Dcal_{\beta}^{(n_{\beta}-1)}(t)}
\dket{\Dcal_{\alpha}^{j_{\alpha}}(t_{0})}\dbra{\Ecal_{\beta}^{\ell_{\beta}}(t_{0})} \text{ , }
\nonumber
\end{align}
where we have used the quasi-eigenvalue relationship, provided by Eq.~(\ref{EqEqEigenStateLa}). Thus, by applying the bi-orthonormality property of the basis, we write
\begin{align}
\Acal_{2}^{\alpha\beta}
&=
e^{\int_{t_{0}}^{t} \left[ \lambda_{\beta}(\xi)-\lambda_{\alpha}(\xi) \right] d\xi}
\sum _{j_{\alpha} = 1}^{N_{\alpha}} \sum _{k_{\alpha} = 1}^{N_{\alpha}}
\sum _{n_{\beta} = 1}^{N_{\beta}} \sum _{\ell_{\beta} = 1}^{N_{\beta}}\left[
\tilde{v}_{j_{\alpha}k_{\alpha}} v_{n_{\beta}\ell_{\beta}}\lambda_{\alpha}(t)
\delta_{\alpha\beta}\delta_{k_{\alpha}n_{\beta}}
\dket{\Dcal_{\alpha}^{j_{\alpha}}(t_{0})}\dbra{\Ecal_{\beta}^{\ell_{\beta}}(t_{0})}
+
\tilde{v}_{j_{\alpha}k_{\alpha}} v_{n_{\beta}\ell_{\beta}}
\delta_{\alpha \beta} \delta_{k_{\alpha}(n_{\beta}-1)}
\dket{\Dcal_{\alpha}^{j_{\alpha}}(t_{0})}\dbra{\Ecal_{\beta}^{\ell_{\beta}}(t_{0})}\right] .
\end{align}
Due to the Kronecker delta $\delta_{\alpha \beta}$, we then write
\begin{align}
\Acal_{2} &= 
\sum_{\alpha = 0}^{N-1}
\sum _{j_{\alpha} = 1}^{N_{\alpha}} \sum _{k_{\alpha} = 1}^{N_{\alpha}}
\sum _{n_{\alpha} = 1}^{N_{\alpha}} \sum _{\ell_{\alpha} = 1}^{N_{\alpha}}
\tilde{v}_{j_{\alpha}k_{\alpha}} v_{n_{\alpha}\ell_{\alpha}}\lambda_{\alpha}(t)
\delta_{k_{\alpha}n_{\alpha}}
\dket{\Dcal_{\alpha}^{j_{\alpha}}(t_{0})}\dbra{\Ecal_{\alpha}^{\ell_{\alpha}}(t_{0})}
+
\sum_{\alpha = 0}^{N-1}
\sum _{j_{\alpha} = 1}^{N_{\alpha}} \sum _{k_{\alpha} = 1}^{N_{\alpha}}
\sum _{n_{\alpha} = 1}^{N_{\alpha}} \sum _{\ell_{\alpha} = 1}^{N_{\alpha}}
\tilde{v}_{j_{\alpha}k_{\alpha}} v_{n_{\alpha}\ell_{\alpha}}
\delta_{k_{\alpha}(n_{\alpha}-1)}
\dket{\Dcal_{\alpha}^{j_{\alpha}}(t_{0})}\dbra{\Ecal_{\alpha}^{\ell_{\alpha}}(t_{0})}
\nonumber\\
&= 
\sum_{\alpha = 0}^{N-1}
\sum _{j_{\alpha} = 1}^{N_{\alpha}}
\sum _{n_{\alpha} = 1}^{N_{\alpha}} \sum _{\ell_{\alpha} = 1}^{N_{\alpha}} \left( \tilde{v}_{j_{\alpha}n_{\alpha}} v_{n_{\alpha}\ell_{\alpha}}\lambda_{\alpha}(t)
+ \tilde{v}_{j_{\alpha}(n_{\alpha}-1)} v_{n_{\alpha}\ell_{\alpha}}\right)
\dket{\Dcal_{\alpha}^{j_{\alpha}}(t_{0})}\dbra{\Ecal_{\alpha}^{\ell_{\alpha}}(t_{0})} \text{ . }
\end{align}
By computing the matrix elements of $\Acal_{2}$ in the right $\{ \dket{\Dcal_{\alpha}^{j_{\alpha}}(t_{0})} \}$ and left $\{\dbra{\Ecal_{\alpha}^{\ell_{\alpha}}(t_{0})} \}$ bases, we get
\begin{align}
\dbra{\Ecal_{\eta}^{g_{\eta}}(t_{0})}\Acal_{2}\dket{\Dcal_{\nu}^{l_{\nu}}(t_{0})} &= 
\sum_{\alpha = 0}^{N-1}
\sum _{j_{\alpha} = 1}^{N_{\alpha}}
\sum _{n_{\alpha} = 1}^{N_{\alpha}} \sum _{\ell_{\alpha} = 1}^{N_{\alpha}} \left( \tilde{v}_{j_{\alpha}n_{\alpha}} v_{n_{\alpha}\ell_{\alpha}}\lambda_{\alpha}(t)
+ \tilde{v}_{j_{\alpha}(n_{\alpha}-1)} v_{n_{\alpha}\ell_{\alpha}}\right)
\delta_{\eta\alpha}\delta_{g_{\eta}j_{\alpha}}
\delta_{\alpha\nu}\delta_{\ell_{\alpha}l_{\nu}} = 
\sum _{n_{\eta} = 1}^{N_{\eta}} \left( \tilde{v}_{g_{\eta}n_{\eta}} v_{n_{\eta}l_{\nu}}\lambda_{\eta}(t)
+ \tilde{v}_{g_{\eta}(n_{\eta}-1)} v_{n_{\eta}l_{\nu}}\right)
\delta_{\eta\nu} .
\end{align}
\end{widetext}
As a first result, we can see that $\dbra{\Ecal_{\eta}^{g_{\eta}}(t_{0})}\Acal_{2}\dket{\Dcal_{\nu}^{l_{\nu}}(t_{0})}$ is nonvanishing only for basis vectors 
$\{\dbra{\Ecal_{\alpha}^{\ell_{\alpha}}(t_{0})} \}$ and $\{ \dket{\Dcal_{\alpha}^{j_{\alpha}}(t_{0})} \}$ 
belonging to the same Jordan block, which means that $\Acal_{2}$ is block diagonal in this basis. Then, for matrix elements inside a Jordan block, we write
\begin{align}
\dbra{\Ecal_{\nu}^{g_{\nu}}(t_{0})}\Acal_{2}\dket{\Dcal_{\nu}^{l_{\nu}}(t_{0})}
&= \lambda_{\nu}(t)
\sum _{n_{\nu} = 1}^{N_{\nu}} \tilde{v}_{g_{\nu}n_{\nu}} v_{n_{\nu}l_{\nu}}
+ \sum _{n_{\nu} = 1}^{N_{\nu}} \tilde{v}_{g_{\nu}(n_{\nu}-1)} v_{n_{\nu}l_{\nu}} \text{ . }
\end{align}
The Jordan decomposition for $\Lmath(t)$ is then achieved both by imposing Eq.~(\ref{ApEqmuINV}) and by requiring
\begin{equation}
\sum _{n_{\nu} = 1}^{N_{\nu}} \tilde{v}_{g_{\nu}(n_{\nu}-1)}(t) v_{n_{\nu}l_{\nu}}(t) = \delta_{l_\nu\, (g_\nu+1)} , \label{3rdcondU}
\end{equation}
with $\tilde{v}_{{g_{\nu}0}}\equiv 0$.  Eq.~(\ref{3rdcondU}) ensures that the neighboring elements of the main diagonal are set to $1$, as required by the Jordan form. 
This equation is automatically satisfied for one dimensional Jordan blocks, but it is nontrivial in the multidimensional case. 


\begin{thebibliography}{41}%
	\makeatletter
	\providecommand \@ifxundefined [1]{%
		\@ifx{#1\undefined}
	}%
	\providecommand \@ifnum [1]{%
		\ifnum #1\expandafter \@firstoftwo
		\else \expandafter \@secondoftwo
		\fi
	}%
	\providecommand \@ifx [1]{%
		\ifx #1\expandafter \@firstoftwo
		\else \expandafter \@secondoftwo
		\fi
	}%
	\providecommand \natexlab [1]{#1}%
	\providecommand \enquote  [1]{``#1''}%
	\providecommand \bibnamefont  [1]{#1}%
	\providecommand \bibfnamefont [1]{#1}%
	\providecommand \citenamefont [1]{#1}%
	\providecommand \href@noop [0]{\@secondoftwo}%
	\providecommand \href [0]{\begingroup \@sanitize@url \@href}%
	\providecommand \@href[1]{\@@startlink{#1}\@@href}%
	\providecommand \@@href[1]{\endgroup#1\@@endlink}%
	\providecommand \@sanitize@url [0]{\catcode `\\12\catcode `\$12\catcode
		`\&12\catcode `\#12\catcode `\^12\catcode `\_12\catcode `\%12\relax}%
	\providecommand \@@startlink[1]{}%
	\providecommand \@@endlink[0]{}%
	\providecommand \url  [0]{\begingroup\@sanitize@url \@url }%
	\providecommand \@url [1]{\endgroup\@href {#1}{\urlprefix }}%
	\providecommand \urlprefix  [0]{URL }%
	\providecommand \Eprint [0]{\href }%
	\providecommand \doibase [0]{http://dx.doi.org/}%
	\providecommand \selectlanguage [0]{\@gobble}%
	\providecommand \bibinfo  [0]{\@secondoftwo}%
	\providecommand \bibfield  [0]{\@secondoftwo}%
	\providecommand \translation [1]{[#1]}%
	\providecommand \BibitemOpen [0]{}%
	\providecommand \bibitemStop [0]{}%
	\providecommand \bibitemNoStop [0]{.\EOS\space}%
	\providecommand \EOS [0]{\spacefactor3000\relax}%
	\providecommand \BibitemShut  [1]{\csname bibitem#1\endcsname}%
	\let\auto@bib@innerbib\@empty
	\bibitem [{\citenamefont {Chen}\ \emph {et~al.}(2011)\citenamefont {Chen},
		\citenamefont {Torrontegui},\ and\ \citenamefont {Muga}}]{Chen:11}%
	\BibitemOpen
	\bibfield  {author} {\bibinfo {author} {\bibfnamefont {X.}~\bibnamefont
			{Chen}}, \bibinfo {author} {\bibfnamefont {E.}~\bibnamefont {Torrontegui}}, \
		and\ \bibinfo {author} {\bibfnamefont {J.~G.}\ \bibnamefont {Muga}},\ }\href
	{\doibase 10.1103/PhysRevA.83.062116} {\bibfield  {journal} {\bibinfo
			{journal} {Phys. Rev. A}\ }\textbf {\bibinfo {volume} {83}},\ \bibinfo
		{pages} {062116} (\bibinfo {year} {2011})}\BibitemShut {NoStop}%
	\bibitem [{\citenamefont {Jing}\ \emph {et~al.}(2013)\citenamefont {Jing},
		\citenamefont {Wu}, \citenamefont {Sarandy},\ and\ \citenamefont
		{Muga}}]{Jing:13}%
	\BibitemOpen
	\bibfield  {author} {\bibinfo {author} {\bibfnamefont {J.}~\bibnamefont
			{Jing}}, \bibinfo {author} {\bibfnamefont {L.-A.}\ \bibnamefont {Wu}},
		\bibinfo {author} {\bibfnamefont {M.~S.}\ \bibnamefont {Sarandy}}, \ and\
		\bibinfo {author} {\bibfnamefont {J.~G.}\ \bibnamefont {Muga}},\ }\href
	{\doibase 10.1103/PhysRevA.88.022333} {\bibfield  {journal} {\bibinfo
			{journal} {Phys. Rev. A}\ }\textbf {\bibinfo {volume} {88}},\ \bibinfo
		{pages} {053422} (\bibinfo {year} {2013})}\BibitemShut {NoStop}%
	\bibitem [{\citenamefont {Kang}\ \emph {et~al.}(2016)\citenamefont {Kang},
		\citenamefont {Chen}, \citenamefont {Wu}, \citenamefont {Huang},
		\citenamefont {Xia},\ and\ \citenamefont {Song}}]{Kang:16}%
	\BibitemOpen
	\bibfield  {author} {\bibinfo {author} {\bibfnamefont {Y.-H.}\ \bibnamefont
			{Kang}}, \bibinfo {author} {\bibfnamefont {Y.-H.}\ \bibnamefont {Chen}},
		\bibinfo {author} {\bibfnamefont {Q.-C.}\ \bibnamefont {Wu}}, \bibinfo
		{author} {\bibfnamefont {B.-H.}\ \bibnamefont {Huang}}, \bibinfo {author}
		{\bibfnamefont {Y.}~\bibnamefont {Xia}}, \ and\ \bibinfo {author}
		{\bibfnamefont {J.}~\bibnamefont {Song}},\ }\href {\doibase
		10.1038/srep30151} {\bibfield  {journal} {\bibinfo  {journal} {Sci. Rep.}\
		}\textbf {\bibinfo {volume} {6}},\ \bibinfo {pages} {30151} (\bibinfo {year}
		{2016})}\BibitemShut {NoStop}%
	\bibitem [{\citenamefont {Yu}\ \emph {et~al.}(2018)\citenamefont {Yu},
		\citenamefont {Zhang}, \citenamefont {Ban},\ and\ \citenamefont
		{Chen}}]{Yu:18}%
	\BibitemOpen
	\bibfield  {author} {\bibinfo {author} {\bibfnamefont {X.-T.}\ \bibnamefont
			{Yu}}, \bibinfo {author} {\bibfnamefont {Q.}~\bibnamefont {Zhang}}, \bibinfo
		{author} {\bibfnamefont {Y.}~\bibnamefont {Ban}}, \ and\ \bibinfo {author}
		{\bibfnamefont {X.}~\bibnamefont {Chen}},\ }\href {\doibase
		10.1103/PhysRevA.97.062317} {\bibfield  {journal} {\bibinfo  {journal} {Phys.
				Rev. A}\ }\textbf {\bibinfo {volume} {97}},\ \bibinfo {pages} {062317}
		(\bibinfo {year} {2018})}\BibitemShut {NoStop}%
	\bibitem [{\citenamefont {Santos}(2018)}]{Santos:18-a}%
	\BibitemOpen
	\bibfield  {author} {\bibinfo {author} {\bibfnamefont {A.~C.}\ \bibnamefont
			{Santos}},\ }\href {\doibase 10.1088/1361-6455/aa987c} {\bibfield  {journal}
		{\bibinfo  {journal} {J. Phys. B: At. Mol. Opt. Phys.}\ }\textbf {\bibinfo
			{volume} {51}},\ \bibinfo {pages} {015501} (\bibinfo {year}
		{2018})}\BibitemShut {NoStop}%
	\bibitem [{\citenamefont {Chen}\ \emph {et~al.}(2018)\citenamefont {Chen},
		\citenamefont {Shi}, \citenamefont {Song},\ and\ \citenamefont
		{Xia}}]{Chen:18-1}%
	\BibitemOpen
	\bibfield  {author} {\bibinfo {author} {\bibfnamefont {Y.-H.}\ \bibnamefont
			{Chen}}, \bibinfo {author} {\bibfnamefont {Z.-C.}\ \bibnamefont {Shi}},
		\bibinfo {author} {\bibfnamefont {J.}~\bibnamefont {Song}}, \ and\ \bibinfo
		{author} {\bibfnamefont {Y.}~\bibnamefont {Xia}},\ }\href {\doibase
		10.1103/PhysRevA.97.023841} {\bibfield  {journal} {\bibinfo  {journal} {Phys.
				Rev. A}\ }\textbf {\bibinfo {volume} {97}},\ \bibinfo {pages} {023841}
		(\bibinfo {year} {2018})}\BibitemShut {NoStop}%
	\bibitem [{\citenamefont {Born}\ and\ \citenamefont {Fock}(1928)}]{Born:28}%
	\BibitemOpen
	\bibfield  {author} {\bibinfo {author} {\bibfnamefont {M.}~\bibnamefont
			{Born}}\ and\ \bibinfo {author} {\bibfnamefont {V.}~\bibnamefont {Fock}},\
	}\href {\doibase 10.1007/BF01343193} {\bibfield  {journal} {\bibinfo
			{journal} {Zeitschrift f{\"u}r Physik A Hadrons and Nuclei}\ }\textbf
		{\bibinfo {volume} {51}},\ \bibinfo {pages} {165} (\bibinfo {year}
		{1928})}\BibitemShut {NoStop}%
	\bibitem [{\citenamefont {Messiah}(1962)}]{Messiah:Book}%
	\BibitemOpen
	\bibfield  {author} {\bibinfo {author} {\bibfnamefont {A.}~\bibnamefont
			{Messiah}},\ }\href {https://archive.org/details/QuantumMechanicsVolumeI}
	{\emph {\bibinfo {title} {Quantum Mechanics}}},\ Quantum Mechanics\ (\bibinfo
	{publisher} {North-Holland Publishing Company},\ \bibinfo {year}
	{1962})\BibitemShut {NoStop}%
	\bibitem [{\citenamefont {Kieu}(2004)}]{Kieu:04}%
	\BibitemOpen
	\bibfield  {author} {\bibinfo {author} {\bibfnamefont {T.~D.}\ \bibnamefont
			{Kieu}},\ }\href {\doibase 10.1103/PhysRevLett.93.140403} {\bibfield
		{journal} {\bibinfo  {journal} {Phys. Rev. Lett.}\ }\textbf {\bibinfo
			{volume} {93}},\ \bibinfo {pages} {140403} (\bibinfo {year}
		{2004})}\BibitemShut {NoStop}%
	\bibitem [{\citenamefont {Alicki}\ and\ \citenamefont
		{Kosloff}(2018)}]{Alicki:18}%
	\BibitemOpen
	\bibfield  {author} {\bibinfo {author} {\bibfnamefont {R.}~\bibnamefont
			{Alicki}}\ and\ \bibinfo {author} {\bibfnamefont {R.}~\bibnamefont
			{Kosloff}},\ }\enquote {\bibinfo {title} {Introduction to quantum
			thermodynamics: History and prospects},}\ in\ \href {\doibase
		10.1007/978-3-319-99046-0_1} {\emph {\bibinfo {booktitle} {Thermodynamics in
				the Quantum Regime: Fundamental Aspects and New Directions}}},\ \bibinfo
	{editor} {edited by\ \bibinfo {editor} {\bibfnamefont {F.}~\bibnamefont
			{Binder}}, \bibinfo {editor} {\bibfnamefont {L.~A.}\ \bibnamefont {Correa}},
		\bibinfo {editor} {\bibfnamefont {C.}~\bibnamefont {Gogolin}}, \bibinfo
		{editor} {\bibfnamefont {J.}~\bibnamefont {Anders}}, \ and\ \bibinfo {editor}
		{\bibfnamefont {G.}~\bibnamefont {Adesso}}}\ (\bibinfo  {publisher} {Springer
		International Publishing},\ \bibinfo {address} {Cham},\ \bibinfo {year}
	{2018})\ pp.\ \bibinfo {pages} {1--33}\BibitemShut {NoStop}%
	\bibitem [{\citenamefont {Hu}\ \emph {et~al.}(2020)\citenamefont {Hu},
		\citenamefont {Santos}, \citenamefont {Cui}, \citenamefont {Huang},
		\citenamefont {Soares-Pinto}, \citenamefont {Sarandy}, \citenamefont {Li},\
		and\ \citenamefont {Guo}}]{Hu:20-a}%
	\BibitemOpen
	\bibfield  {author} {\bibinfo {author} {\bibfnamefont {C.-K.}\ \bibnamefont
			{Hu}}, \bibinfo {author} {\bibfnamefont {A.~C.}\ \bibnamefont {Santos}},
		\bibinfo {author} {\bibfnamefont {J.-M.}\ \bibnamefont {Cui}}, \bibinfo
		{author} {\bibfnamefont {Y.-F.}\ \bibnamefont {Huang}}, \bibinfo {author}
		{\bibfnamefont {D.~O.}\ \bibnamefont {Soares-Pinto}}, \bibinfo {author}
		{\bibfnamefont {M.~S.}\ \bibnamefont {Sarandy}}, \bibinfo {author}
		{\bibfnamefont {C.-F.}\ \bibnamefont {Li}}, \ and\ \bibinfo {author}
		{\bibfnamefont {G.-C.}\ \bibnamefont {Guo}},\ }\href {\doibase
		10.1038/s41534-020-00300-2} {\bibfield  {journal} {\bibinfo  {journal} {npj
				Quantum Information}\ }\textbf {\bibinfo {volume} {6}},\ \bibinfo {pages}
		{73} (\bibinfo {year} {2020})}\BibitemShut {NoStop}%
	\bibitem [{\citenamefont {Deffner}\ and\ \citenamefont
		{Campbell}(2019)}]{Deffner-Campbel:Book}%
	\BibitemOpen
	\bibfield  {author} {\bibinfo {author} {\bibfnamefont {S.}~\bibnamefont
			{Deffner}}\ and\ \bibinfo {author} {\bibfnamefont {S.}~\bibnamefont
			{Campbell}},\ }\href {\doibase 10.1088/2053-2571/ab21c6} {\emph {\bibinfo
			{title} {Quantum Thermodynamics}}}\ (\bibinfo  {publisher} {Morgan \&
		Claypool Publishers},\ \bibinfo {year} {2019})\BibitemShut {NoStop}%
	\bibitem [{\citenamefont {Kr\'al}\ \emph {et~al.}(2007)\citenamefont {Kr\'al},
		\citenamefont {Thanopulos},\ and\ \citenamefont {Shapiro}}]{Kral:07}%
	\BibitemOpen
	\bibfield  {author} {\bibinfo {author} {\bibfnamefont {P.}~\bibnamefont
			{Kr\'al}}, \bibinfo {author} {\bibfnamefont {I.}~\bibnamefont {Thanopulos}},
		\ and\ \bibinfo {author} {\bibfnamefont {M.}~\bibnamefont {Shapiro}},\ }\href
	{\doibase 10.1103/RevModPhys.79.53} {\bibfield  {journal} {\bibinfo
			{journal} {Rev. Mod. Phys.}\ }\textbf {\bibinfo {volume} {79}},\ \bibinfo
		{pages} {53} (\bibinfo {year} {2007})}\BibitemShut {NoStop}%
	\bibitem [{\citenamefont {Gu\'ery-Odelin}\ \emph {et~al.}(2019)\citenamefont
		{Gu\'ery-Odelin}, \citenamefont {Ruschhaupt}, \citenamefont {Kiely},
		\citenamefont {Torrontegui}, \citenamefont {Mart\'{\i}nez-Garaot},\ and\
		\citenamefont {Muga}}]{Odelin:19}%
	\BibitemOpen
	\bibfield  {author} {\bibinfo {author} {\bibfnamefont {D.}~\bibnamefont
			{Gu\'ery-Odelin}}, \bibinfo {author} {\bibfnamefont {A.}~\bibnamefont
			{Ruschhaupt}}, \bibinfo {author} {\bibfnamefont {A.}~\bibnamefont {Kiely}},
		\bibinfo {author} {\bibfnamefont {E.}~\bibnamefont {Torrontegui}}, \bibinfo
		{author} {\bibfnamefont {S.}~\bibnamefont {Mart\'{\i}nez-Garaot}}, \ and\
		\bibinfo {author} {\bibfnamefont {J.~G.}\ \bibnamefont {Muga}},\ }\href
	{\doibase 10.1103/RevModPhys.91.045001} {\bibfield  {journal} {\bibinfo
			{journal} {Rev. Mod. Phys.}\ }\textbf {\bibinfo {volume} {91}},\ \bibinfo
		{pages} {045001} (\bibinfo {year} {2019})}\BibitemShut {NoStop}%
	\bibitem [{\citenamefont {Farhi}\ \emph {et~al.}(2001)\citenamefont {Farhi},
		\citenamefont {Goldstone}, \citenamefont {Gutmann}, \citenamefont {Lapan},
		\citenamefont {Lundgren},\ and\ \citenamefont {Preda}}]{Farhi:01}%
	\BibitemOpen
	\bibfield  {author} {\bibinfo {author} {\bibfnamefont {E.}~\bibnamefont
			{Farhi}}, \bibinfo {author} {\bibfnamefont {J.}~\bibnamefont {Goldstone}},
		\bibinfo {author} {\bibfnamefont {S.}~\bibnamefont {Gutmann}}, \bibinfo
		{author} {\bibfnamefont {J.}~\bibnamefont {Lapan}}, \bibinfo {author}
		{\bibfnamefont {A.}~\bibnamefont {Lundgren}}, \ and\ \bibinfo {author}
		{\bibfnamefont {D.}~\bibnamefont {Preda}},\ }\href {\doibase
		10.1126/science.1057726} {\bibfield  {journal} {\bibinfo  {journal}
			{Science}\ }\textbf {\bibinfo {volume} {292}},\ \bibinfo {pages} {472}
		(\bibinfo {year} {2001})}\BibitemShut {NoStop}%
	\bibitem [{\citenamefont {Albash}\ and\ \citenamefont
		{Lidar}(2018)}]{Tameem:18}%
	\BibitemOpen
	\bibfield  {author} {\bibinfo {author} {\bibfnamefont {T.}~\bibnamefont
			{Albash}}\ and\ \bibinfo {author} {\bibfnamefont {D.~A.}\ \bibnamefont
			{Lidar}},\ }\href {\doibase 10.1103/RevModPhys.90.015002} {\bibfield
		{journal} {\bibinfo  {journal} {Rev. Mod. Phys.}\ }\textbf {\bibinfo {volume}
			{90}},\ \bibinfo {pages} {015002} (\bibinfo {year} {2018})}\BibitemShut
	{NoStop}%
	\bibitem [{\citenamefont {Sarandy}\ and\ \citenamefont
		{Lidar}(2005{\natexlab{a}})}]{Sarandy:05-1}%
	\BibitemOpen
	\bibfield  {author} {\bibinfo {author} {\bibfnamefont {M.~S.}\ \bibnamefont
			{Sarandy}}\ and\ \bibinfo {author} {\bibfnamefont {D.~A.}\ \bibnamefont
			{Lidar}},\ }\href {\doibase 10.1103/PhysRevA.71.012331} {\bibfield  {journal}
		{\bibinfo  {journal} {Phys. Rev. A}\ }\textbf {\bibinfo {volume} {71}},\
		\bibinfo {pages} {012331} (\bibinfo {year} {2005}{\natexlab{a}})}\BibitemShut
	{NoStop}%
	\bibitem [{\citenamefont {Sarandy}\ and\ \citenamefont
		{Lidar}(2005{\natexlab{b}})}]{Sarandy:05-2}%
	\BibitemOpen
	\bibfield  {author} {\bibinfo {author} {\bibfnamefont {M.~S.}\ \bibnamefont
			{Sarandy}}\ and\ \bibinfo {author} {\bibfnamefont {D.~A.}\ \bibnamefont
			{Lidar}},\ }\href {\doibase 10.1103/PhysRevLett.95.250503} {\bibfield
		{journal} {\bibinfo  {journal} {Phys. Rev. Lett.}\ }\textbf {\bibinfo
			{volume} {95}},\ \bibinfo {pages} {250503} (\bibinfo {year}
		{2005}{\natexlab{b}})}\BibitemShut {NoStop}%
	\bibitem [{\citenamefont {Sarandy}\ and\ \citenamefont
		{Lidar}(2006)}]{sarandy:06}%
	\BibitemOpen
	\bibfield  {author} {\bibinfo {author} {\bibfnamefont {M.~S.}\ \bibnamefont
			{Sarandy}}\ and\ \bibinfo {author} {\bibfnamefont {D.~A.}\ \bibnamefont
			{Lidar}},\ }\href {\doibase 10.1103/PhysRevA.73.062101} {\bibfield  {journal}
		{\bibinfo  {journal} {Phys. Rev. A}\ }\textbf {\bibinfo {volume} {73}},\
		\bibinfo {pages} {062101} (\bibinfo {year} {2006})}\BibitemShut {NoStop}%
	\bibitem [{\citenamefont {Jing}\ \emph {et~al.}(2016)\citenamefont {Jing},
		\citenamefont {Sarandy}, \citenamefont {Lidar}, \citenamefont {Luo},\ and\
		\citenamefont {Wu}}]{Jing:16}%
	\BibitemOpen
	\bibfield  {author} {\bibinfo {author} {\bibfnamefont {J.}~\bibnamefont
			{Jing}}, \bibinfo {author} {\bibfnamefont {M.~S.}\ \bibnamefont {Sarandy}},
		\bibinfo {author} {\bibfnamefont {D.~A.}\ \bibnamefont {Lidar}}, \bibinfo
		{author} {\bibfnamefont {D.-W.}\ \bibnamefont {Luo}}, \ and\ \bibinfo
		{author} {\bibfnamefont {L.-A.}\ \bibnamefont {Wu}},\ }\href {\doibase
		10.1103/PhysRevA.94.042131} {\bibfield  {journal} {\bibinfo  {journal} {Phys.
				Rev. A}\ }\textbf {\bibinfo {volume} {94}},\ \bibinfo {pages} {042131}
		(\bibinfo {year} {2016})}\BibitemShut {NoStop}%
	\bibitem [{\citenamefont {Steffen}\ \emph {et~al.}(2003)\citenamefont
		{Steffen}, \citenamefont {van Dam}, \citenamefont {Hogg}, \citenamefont
		{Breyta},\ and\ \citenamefont {Chuang}}]{Steffen:03}%
	\BibitemOpen
	\bibfield  {author} {\bibinfo {author} {\bibfnamefont {M.}~\bibnamefont
			{Steffen}}, \bibinfo {author} {\bibfnamefont {W.}~\bibnamefont {van Dam}},
		\bibinfo {author} {\bibfnamefont {T.}~\bibnamefont {Hogg}}, \bibinfo {author}
		{\bibfnamefont {G.}~\bibnamefont {Breyta}}, \ and\ \bibinfo {author}
		{\bibfnamefont {I.}~\bibnamefont {Chuang}},\ }\href {\doibase
		10.1103/PhysRevLett.90.067903} {\bibfield  {journal} {\bibinfo  {journal}
			{Phys. Rev. Lett.}\ }\textbf {\bibinfo {volume} {90}},\ \bibinfo {pages}
		{067903} (\bibinfo {year} {2003})}\BibitemShut {NoStop}%
	\bibitem [{\citenamefont {Yi}\ \emph {et~al.}(2007)\citenamefont {Yi},
		\citenamefont {Tong}, \citenamefont {Kwek},\ and\ \citenamefont
		{Oh}}]{Yi:07}%
	\BibitemOpen
	\bibfield  {author} {\bibinfo {author} {\bibfnamefont {X.~X.}\ \bibnamefont
			{Yi}}, \bibinfo {author} {\bibfnamefont {D.~M.}\ \bibnamefont {Tong}},
		\bibinfo {author} {\bibfnamefont {L.~C.}\ \bibnamefont {Kwek}}, \ and\
		\bibinfo {author} {\bibfnamefont {C.~H.}\ \bibnamefont {Oh}},\ }\href
	{\doibase 10.1088/0953-4075/40/2/004} {\bibfield  {journal} {\bibinfo
			{journal} {J. Phys. B: At. Mol. Opt. Phys.}\ }\textbf {\bibinfo {volume}
			{40}},\ \bibinfo {pages} {281} (\bibinfo {year} {2007})}\BibitemShut
	{NoStop}%
	\bibitem [{\citenamefont {Oreshkov}\ and\ \citenamefont
		{Calsamiglia}(2010)}]{Oreshkov:10}%
	\BibitemOpen
	\bibfield  {author} {\bibinfo {author} {\bibfnamefont {O.}~\bibnamefont
			{Oreshkov}}\ and\ \bibinfo {author} {\bibfnamefont {J.}~\bibnamefont
			{Calsamiglia}},\ }\href {\doibase 10.1103/PhysRevLett.105.050503} {\bibfield
		{journal} {\bibinfo  {journal} {Phys. Rev. Lett.}\ }\textbf {\bibinfo
			{volume} {105}},\ \bibinfo {pages} {050503} (\bibinfo {year}
		{2010})}\BibitemShut {NoStop}%
	\bibitem [{\citenamefont {Thunstr\"om}\ \emph {et~al.}(2005)\citenamefont
		{Thunstr\"om}, \citenamefont {\AA{}berg},\ and\ \citenamefont
		{Sj\"oqvist}}]{Patrik:05}%
	\BibitemOpen
	\bibfield  {author} {\bibinfo {author} {\bibfnamefont {P.}~\bibnamefont
			{Thunstr\"om}}, \bibinfo {author} {\bibfnamefont {J.}~\bibnamefont
			{\AA{}berg}}, \ and\ \bibinfo {author} {\bibfnamefont {E.}~\bibnamefont
			{Sj\"oqvist}},\ }\href {\doibase 10.1103/PhysRevA.72.022328} {\bibfield
		{journal} {\bibinfo  {journal} {Phys. Rev. A}\ }\textbf {\bibinfo {volume}
			{72}},\ \bibinfo {pages} {022328} (\bibinfo {year} {2005})}\BibitemShut
	{NoStop}%
	\bibitem [{\citenamefont {Venuti}\ \emph {et~al.}(2016)\citenamefont {Venuti},
		\citenamefont {Albash}, \citenamefont {Lidar},\ and\ \citenamefont
		{Zanardi}}]{Venuti:16}%
	\BibitemOpen
	\bibfield  {author} {\bibinfo {author} {\bibfnamefont {L.~C.}\ \bibnamefont
			{Venuti}}, \bibinfo {author} {\bibfnamefont {T.}~\bibnamefont {Albash}},
		\bibinfo {author} {\bibfnamefont {D.~A.}\ \bibnamefont {Lidar}}, \ and\
		\bibinfo {author} {\bibfnamefont {P.}~\bibnamefont {Zanardi}},\ }\href
	{\doibase 10.1103/PhysRevA.93.032118} {\bibfield  {journal} {\bibinfo
			{journal} {Phys. Rev. A}\ }\textbf {\bibinfo {volume} {93}},\ \bibinfo
		{pages} {032118} (\bibinfo {year} {2016})}\BibitemShut {NoStop}%
	\bibitem [{\citenamefont {Avron}\ \emph {et~al.}(2012)\citenamefont {Avron},
		\citenamefont {Fraas}, \citenamefont {Graf},\ and\ \citenamefont
		{Grech}}]{Avron:12}%
	\BibitemOpen
	\bibfield  {author} {\bibinfo {author} {\bibfnamefont {J.~E.}\ \bibnamefont
			{Avron}}, \bibinfo {author} {\bibfnamefont {M.}~\bibnamefont {Fraas}},
		\bibinfo {author} {\bibfnamefont {G.~M.}\ \bibnamefont {Graf}}, \ and\
		\bibinfo {author} {\bibfnamefont {P.}~\bibnamefont {Grech}},\ }\href
	{\doibase 10.1007/s00220-012-1504-1} {\bibfield  {journal} {\bibinfo
			{journal} {Commun. Math. Phys.}\ }\textbf {\bibinfo {volume} {314}},\
		\bibinfo {pages} {163} (\bibinfo {year} {2012})}\BibitemShut {NoStop}%
	\bibitem [{\citenamefont {Tong}\ \emph {et~al.}(2007)\citenamefont {Tong},
		\citenamefont {Singh}, \citenamefont {Kwek},\ and\ \citenamefont
		{Oh}}]{Tong:07}%
	\BibitemOpen
	\bibfield  {author} {\bibinfo {author} {\bibfnamefont {D.~M.}\ \bibnamefont
			{Tong}}, \bibinfo {author} {\bibfnamefont {K.}~\bibnamefont {Singh}},
		\bibinfo {author} {\bibfnamefont {L.~C.}\ \bibnamefont {Kwek}}, \ and\
		\bibinfo {author} {\bibfnamefont {C.~H.}\ \bibnamefont {Oh}},\ }\href
	{\doibase 10.1103/PhysRevLett.98.150402} {\bibfield  {journal} {\bibinfo
			{journal} {Phys. Rev. Lett.}\ }\textbf {\bibinfo {volume} {98}},\ \bibinfo
		{pages} {150402} (\bibinfo {year} {2007})}\BibitemShut {NoStop}%
	\bibitem [{\citenamefont {Sarandy}\ \emph {et~al.}(2004)\citenamefont
		{Sarandy}, \citenamefont {Wu},\ and\ \citenamefont {Lidar}}]{Sarandy:04}%
	\BibitemOpen
	\bibfield  {author} {\bibinfo {author} {\bibfnamefont {M.~S.}\ \bibnamefont
			{Sarandy}}, \bibinfo {author} {\bibfnamefont {L.-A.}\ \bibnamefont {Wu}}, \
		and\ \bibinfo {author} {\bibfnamefont {D.~A.}\ \bibnamefont {Lidar}},\ }\href
	{\doibase 10.1007/s11128-004-7712-7} {\bibfield  {journal} {\bibinfo
			{journal} {Quantum Information Processing}\ }\textbf {\bibinfo {volume}
			{3}},\ \bibinfo {pages} {331} (\bibinfo {year} {2004})}\BibitemShut {NoStop}%
	\bibitem [{\citenamefont {Horn}\ and\ \citenamefont
		{Johnson}(2012)}]{Horn:Book}%
	\BibitemOpen
	\bibfield  {author} {\bibinfo {author} {\bibfnamefont {R.~A.}\ \bibnamefont
			{Horn}}\ and\ \bibinfo {author} {\bibfnamefont {C.~R.}\ \bibnamefont
			{Johnson}},\ }\href {\doibase 10.1017/9781139020411} {\emph {\bibinfo {title}
			{Matrix analysis}}},\ \bibinfo {edition} {2nd}\ ed.\ (\bibinfo  {publisher}
	{Cambridge University Press},\ \bibinfo {address} {Cambridge},\ \bibinfo
	{year} {2012})\BibitemShut {NoStop}%
	\bibitem [{\citenamefont {de~Muynck}(2002)}]{Muynck:Book}%
	\BibitemOpen
	\bibfield  {author} {\bibinfo {author} {\bibfnamefont {W.~M.}\ \bibnamefont
			{de~Muynck}},\ }\href {\doibase 10.1007/0-306-48047-6} {\emph {\bibinfo
			{title} {Foundations of Quantum Mechanics: an Empiricist Approach}}},\
	\bibinfo {edition} {1st}\ ed.\ (\bibinfo  {publisher} {Kluwer Academic
		Publishers},\ \bibinfo {address} {Dordrecht, Netherlands},\ \bibinfo {year}
	{2002})\BibitemShut {NoStop}%
	\bibitem [{\citenamefont {{Santos}}\ \emph {et~al.}(2020)\citenamefont
		{{Santos}}, \citenamefont {{Villas-Boas}},\ and\ \citenamefont
		{{Bachelard}}}]{Santos:20c}%
	\BibitemOpen
	\bibfield  {author} {\bibinfo {author} {\bibfnamefont {A.~C.}\ \bibnamefont
			{{Santos}}}, \bibinfo {author} {\bibfnamefont {C.~J.}\ \bibnamefont
			{{Villas-Boas}}}, \ and\ \bibinfo {author} {\bibfnamefont {R.}~\bibnamefont
			{{Bachelard}}},\ }\href@noop {} {\bibfield  {journal} {\bibinfo  {journal}
			{arXiv e-prints}\ ,\ \bibinfo {eid} {arXiv:2006.13718}} (\bibinfo {year}
		{2020})},\ \Eprint {http://arxiv.org/abs/2006.13718} {arXiv:2006.13718
		[quant-ph]} \BibitemShut {NoStop}%
	\bibitem [{\citenamefont {Hu}\ \emph {et~al.}(2019)\citenamefont {Hu},
		\citenamefont {Santos}, \citenamefont {Cui}, \citenamefont {Huang},
		\citenamefont {Sarandy}, \citenamefont {Li},\ and\ \citenamefont
		{Guo}}]{Hu:19-a}%
	\BibitemOpen
	\bibfield  {author} {\bibinfo {author} {\bibfnamefont {C.-K.}\ \bibnamefont
			{Hu}}, \bibinfo {author} {\bibfnamefont {A.~C.}\ \bibnamefont {Santos}},
		\bibinfo {author} {\bibfnamefont {J.-M.}\ \bibnamefont {Cui}}, \bibinfo
		{author} {\bibfnamefont {Y.-F.}\ \bibnamefont {Huang}}, \bibinfo {author}
		{\bibfnamefont {M.~S.}\ \bibnamefont {Sarandy}}, \bibinfo {author}
		{\bibfnamefont {C.-F.}\ \bibnamefont {Li}}, \ and\ \bibinfo {author}
		{\bibfnamefont {G.-C.}\ \bibnamefont {Guo}},\ }\href {\doibase
		10.1103/PhysRevA.99.062320} {\bibfield  {journal} {\bibinfo  {journal} {Phys.
				Rev. A}\ }\textbf {\bibinfo {volume} {99}},\ \bibinfo {pages} {062320}
		(\bibinfo {year} {2019})}\BibitemShut {NoStop}%
	\bibitem [{\citenamefont {Ib\'a\~nez}\ \emph {et~al.}(2012)\citenamefont
		{Ib\'a\~nez}, \citenamefont {Chen}, \citenamefont {Torrontegui},
		\citenamefont {Muga},\ and\ \citenamefont {Ruschhaupt}}]{Ibanez:12}%
	\BibitemOpen
	\bibfield  {author} {\bibinfo {author} {\bibfnamefont {S.}~\bibnamefont
			{Ib\'a\~nez}}, \bibinfo {author} {\bibfnamefont {X.}~\bibnamefont {Chen}},
		\bibinfo {author} {\bibfnamefont {E.}~\bibnamefont {Torrontegui}}, \bibinfo
		{author} {\bibfnamefont {J.~G.}\ \bibnamefont {Muga}}, \ and\ \bibinfo
		{author} {\bibfnamefont {A.}~\bibnamefont {Ruschhaupt}},\ }\href {\doibase
		10.1103/PhysRevLett.109.100403} {\bibfield  {journal} {\bibinfo  {journal}
			{Phys. Rev. Lett.}\ }\textbf {\bibinfo {volume} {109}},\ \bibinfo {pages}
		{100403} (\bibinfo {year} {2012})}\BibitemShut {NoStop}%
	\bibitem [{\citenamefont {Garrido}(1964)}]{Garrido:64}%
	\BibitemOpen
	\bibfield  {author} {\bibinfo {author} {\bibfnamefont {L.~M.}\ \bibnamefont
			{Garrido}},\ }\href {\doibase 10.1063/1.1704127} {\bibfield  {journal}
		{\bibinfo  {journal} {Journal of Mathematical Physics}\ }\textbf {\bibinfo
			{volume} {5}},\ \bibinfo {pages} {355} (\bibinfo {year} {1964})}\BibitemShut
	{NoStop}%
	\bibitem [{\citenamefont {Berry}(1987)}]{Berry:87}%
	\BibitemOpen
	\bibfield  {author} {\bibinfo {author} {\bibfnamefont {M.~V.}\ \bibnamefont
			{Berry}},\ }\href {\doibase 10.1098/rspa.1987.0131} {\bibfield  {journal}
		{\bibinfo  {journal} {Proceedings of the Royal Society of London. A.
				Mathematical and Physical Sciences}\ }\textbf {\bibinfo {volume} {414}},\
		\bibinfo {pages} {31} (\bibinfo {year} {1987})}\BibitemShut {NoStop}%
	\bibitem [{\citenamefont {Deutsch}(1985)}]{Deutsch:85}%
	\BibitemOpen
	\bibfield  {author} {\bibinfo {author} {\bibfnamefont {D.}~\bibnamefont
			{Deutsch}},\ }\href {\doibase 10.1098/rspa.1985.0070} {\bibfield  {journal}
		{\bibinfo  {journal} {Proc. R. Soc. Lond. A}\ }\textbf {\bibinfo {volume}
			{400}},\ \bibinfo {pages} {97} (\bibinfo {year} {1985})}\BibitemShut
	{NoStop}%
	\bibitem [{\citenamefont {Nielsen}\ and\ \citenamefont
		{Chuang}(2011)}]{Nielsen:Book}%
	\BibitemOpen
	\bibfield  {author} {\bibinfo {author} {\bibfnamefont {M.~A.}\ \bibnamefont
			{Nielsen}}\ and\ \bibinfo {author} {\bibfnamefont {I.~L.}\ \bibnamefont
			{Chuang}},\ }\href {\doibase 10.1017/CBO9780511976667} {\emph {\bibinfo
			{title} {Quantum Computation and Quantum Information: 10th Anniversary
				Edition}}},\ \bibinfo {edition} {10th}\ ed.\ (\bibinfo  {publisher}
	{Cambridge University Press},\ \bibinfo {address} {New York, NY, USA},\
	\bibinfo {year} {2011})\BibitemShut {NoStop}%
	\bibitem [{\citenamefont {Rigolin}\ \emph {et~al.}(2008)\citenamefont
		{Rigolin}, \citenamefont {Ortiz},\ and\ \citenamefont {Ponce}}]{Rigolin:08}%
	\BibitemOpen
	\bibfield  {author} {\bibinfo {author} {\bibfnamefont {G.}~\bibnamefont
			{Rigolin}}, \bibinfo {author} {\bibfnamefont {G.}~\bibnamefont {Ortiz}}, \
		and\ \bibinfo {author} {\bibfnamefont {V.~H.}\ \bibnamefont {Ponce}},\ }\href
	{\doibase 10.1103/PhysRevA.78.052508} {\bibfield  {journal} {\bibinfo
			{journal} {Phys. Rev. A}\ }\textbf {\bibinfo {volume} {78}},\ \bibinfo
		{pages} {052508} (\bibinfo {year} {2008})}\BibitemShut {NoStop}%
	\bibitem [{\citenamefont {Rigolin}\ and\ \citenamefont
		{Ortiz}(2010)}]{Rigolin:10}%
	\BibitemOpen
	\bibfield  {author} {\bibinfo {author} {\bibfnamefont {G.}~\bibnamefont
			{Rigolin}}\ and\ \bibinfo {author} {\bibfnamefont {G.}~\bibnamefont
			{Ortiz}},\ }\href {\doibase 10.1103/PhysRevLett.104.170406} {\bibfield
		{journal} {\bibinfo  {journal} {Phys. Rev. Lett.}\ }\textbf {\bibinfo
			{volume} {104}},\ \bibinfo {pages} {170406} (\bibinfo {year}
		{2010})}\BibitemShut {NoStop}%
	\bibitem [{\citenamefont {Vacanti}\ \emph {et~al.}(2014)\citenamefont
		{Vacanti}, \citenamefont {Fazio}, \citenamefont {Montangero}, \citenamefont
		{Palma}, \citenamefont {Paternostro},\ and\ \citenamefont
		{Vedral}}]{Vacanti:14}%
	\BibitemOpen
	\bibfield  {author} {\bibinfo {author} {\bibfnamefont {G.}~\bibnamefont
			{Vacanti}}, \bibinfo {author} {\bibfnamefont {R.}~\bibnamefont {Fazio}},
		\bibinfo {author} {\bibfnamefont {S.}~\bibnamefont {Montangero}}, \bibinfo
		{author} {\bibfnamefont {G.}~\bibnamefont {Palma}}, \bibinfo {author}
		{\bibfnamefont {M.}~\bibnamefont {Paternostro}}, \ and\ \bibinfo {author}
		{\bibfnamefont {V.}~\bibnamefont {Vedral}},\ }\href {\doibase
		10.1088/1367-2630/16/5/053017} {\bibfield  {journal} {\bibinfo  {journal}
			{New J. Phys.}\ }\textbf {\bibinfo {volume} {16}},\ \bibinfo {pages} {053017}
		(\bibinfo {year} {2014})}\BibitemShut {NoStop}%
	\bibitem [{\citenamefont {Alicki}\ and\ \citenamefont
		{Lendi}(2007)}]{Alicki:Book07}%
	\BibitemOpen
	\bibfield  {author} {\bibinfo {author} {\bibfnamefont {R.}~\bibnamefont
			{Alicki}}\ and\ \bibinfo {author} {\bibfnamefont {K.}~\bibnamefont {Lendi}},\
	}\enquote {\bibinfo {title} {Recent developments},}\ in\ \href {\doibase
		10.1007/3-540-70861-8_3} {\emph {\bibinfo {booktitle} {Quantum Dynamical
				Semigroups and Applications}}}\ (\bibinfo  {publisher} {Springer Berlin
		Heidelberg},\ \bibinfo {address} {Berlin, Heidelberg},\ \bibinfo {year}
	{2007})\ pp.\ \bibinfo {pages} {109--121}\BibitemShut {NoStop}%
\end{thebibliography}

%

\end{document}